\documentclass[onecolumn,aps]{openjournal}
\usepackage{lineno}
\usepackage[colorlinks=True]{hyperref}

\begin{document}

\newcommand\blfootnote[1]{%
  \begingroup
  \renewcommand\thefootnote{}\footnote{#1}%
  \addtocounter{footnote}{-1}%
  \endgroup
}

\newcommand{\fro}{$f_{R_0}$}

\title{Categorizing models using Self-Organizing Maps: an application to modified gravity theories probed by cosmic shear}

    
\author{Agnès Ferté$^{1}$, Shoubaneh Hemmati$^{2}$, Daniel Masters$^{2}$, Brigitte Montminy$^{1}$, Peter L. Taylor$^{1}$, Eric Huff$^{1}$, Jason Rhodes$^{1}$}
\affiliation{
$^1$ Jet Propulsion Laboratory, California Institute of Technology, 4800 Oak Grove Drive, Pasadena, CA, 91109, USA \\
$^2$IPAC, California Institute of Technology, 1200 E California Blvd, Pasadena, CA 91125, USA}
\blfootnote{\copyright 2023. All rights reserved.}

\begin{abstract}
We propose to use Self-Organizing Maps (SOM) to map the impact of physical models onto observables.
Using this approach, we are be able to determine how theories relate to each other given their signatures. 
In cosmology this will be particularly useful to determine cosmological models (such as dark energy, modified gravity or inflationary models) that should be tested by the new generation of experiments.
As a first example, we apply this approach to the representation of a subset of the space of modified gravity theories probed by cosmic shear. 
We therefore train a SOM on shear correlation functions in the $f(R)$, dilaton and symmetron models. The results indicate these three theories have similar signatures on shear for small values of their parameters but the dilaton has different signature for higher values. 
We also show that modified gravity (especially the dilaton model) has a different impact on cosmic shear compared to a dynamical dark energy so both need to be tested by galaxy surveys.
\end{abstract}


\section{Introduction}   

In many fields a large variety of theoretical models are proposed in the literature and are to be tested with current or future data. 
For example, in cosmology 74 inflation models were systematically tested in \cite{martin_2013} using cosmic microwave background 2013 data from the Planck satellite \cite{planck_2013}, or in astrophysics, testing various models of exoplanets' atmosphere will be one of the goals of experiments like the JWST as reviewed by \cite{Madhusudhan_2019,Biller_2018}.
Selecting only a subset of models to be tested given an observable is in some cases necessary as obtaining robust constraints can be time consuming, needs modeling developments specific to each theory or requires large computing power (to run Monte Carlo Markov Chains for instance).
However making a model selection based on the inspection of observables can be inconvenient in the case where models are characterized by several parameters or observables are described by several components, such as tomographic real-space correlation functions in cosmology.
In this paper, we address this challenge by presenting a new approach using unsupervised non-Gaussian dimensionality reduction algorithms to facilitate model selection.  
We especially use Self-Organizing Maps (SOMs) as they can produce 2-dimensional maps where models are assigned to cells preserving topology.
Such mapping indeed allows to easily determine models that have different impacts on a given observable and the effect of measurements on this mapping can also be added (but is not done in the present study as described later on).
This approach complements Principal Component Analysis (PCA) as SOMs are adequate for non-Gaussian reduction and vizualization. 
As a first step and for purpose of example, we apply our approach to the case of modified gravity (MG) theories probed by cosmic shear.
A similar approach has been proposed in \cite{bnn}, using Bayesian Neural Networks as a classifier trained on matter power spectra. In the present study, we instead use a dimensionality reduction algorithm trained on cosmic shear, a probe of the matter power spectrum. 

 Weak gravitational lensing is indeed a powerful probe of gravity as shown early on in \cite{song_looking_2005,ishak_2005} demonstrating that some MG theories can be distinguished from a dark energy using cosmic shear, and later in \cite{song_cosmological_2006,tsujikawa_effect_2008,jain_observational_2008}.
 Tomographic cosmic shear corresponds to the correlation in redshift bins of the deformation of observed galaxy shapes due to weak lensing. It directly probes the matter distribution and is sensitive to the growth of large scale structures in the Universe, which depends on the laws of gravity.
 Deviations from GR were thus tested using cosmic shear data from CFHTLens\footnote{\url{https://www.cfhtlens.org/}} \cite{cfhtlens_fr,simpson_cfhtlens,Dossett_2015,ferte_testing_2017}, KIDS\footnote{\url{http://kids.strw.leidenuniv.nl/}} \cite{joudaki_kids-450_2017,troster_kids-1000_2020} and DES\footnote{\url{https://www.darkenergysurvey.org/}} \cite{des_collaboration_dark_2018} (in the later cases galaxy-galaxy lensing and clustering data were also used).
 Forecasts show these constraints will improve with future imaging surveys including the Euclid satellite\footnote{\url{https://sci.esa.int/web/euclid/}} \cite{martinelli_constraining_2011,euclid_collaboration_euclid_2020}, LSST\footnote{\url{https://www.lsst.org/}} that will be produced by the Rubin Observatory \cite{lsst_forecast,Hojjati_2016,ferte_testing_2017} and the Roman space mission\footnote{\url{https://roman.gsfc.nasa.gov/}} \cite{dore_wfirst_2019,eifler_cosmology_2020,eifler_cosmology_2021}.
 
One of the challenges faced by these future surveys is to decide which theories of MG to test amongst the large space of models proposed in the literature. 
For instance \cite{lsst_mg} prioritizes theories to be tested by LSST according to their theoretical maturity and impact on LSST observables, however without comparing their signatures on observables.
Moreover, for surveys such as LSST, cosmic shear is expected to be measured through correlation functions in 10 redshift bins, corresponding to 55 cross- and auto-correlations, making a visual inspection of the impact of MG models on cosmic shear difficult.
The goal of the present approach is to fill the gap between choosing models to test based on their theoretical interest (such as in \cite{lsst_mg}) and testing a wide variety of theories (e.g. \cite{planck_2015_mg}) or using non-parametric reconstruction \cite{raveri2021joint} or PCA \cite{Linder_2005,pca_sigmamu}): we propose a way to categorize models in two dimensions according to their impact on observables using SOMs.
We do so by producing a training set made of theoretical predictions of cosmic shear in MG models assuming a fixed cosmology. 
We indeed want to know how MG models would compare to each other if we had a perfect knowledge of cosmology.
If two MG models have the same impact at fixed cosmology then there is definitely no chance to distinguish between them when doing a standard cosmological inference analysis, studying one model is therefore sufficient.
So in this paper, we set $\Omega_m $ = 0.31, $H_0$ = 67, $\Omega_b$ = 0.04, $A_s = 2.1 \times 10^{-9}$, $n_s = 0.96$, one massive neutrino with a mass of 0.06 eV. 
The training sets are available on request and the script to reproduce results presented in this paper is available online\footnote{\url{https://github.com/aferte/CosmoSOM-MG}}.

\section{Self-Organizing Map of modified gravity probed by cosmic shear}

\textit{Self-Organizing Maps --} Self-Organizing Maps, introduced in \cite{kohonen_self-organized_1982}, are a class of unsupervised neural networks which learn and reduce dimensions (usually to 2 dimensions) of a higher dimensional data set while preserving the topology of the manifold. 
Dimensionality reduction makes SOMs perfect tools for visualization. 
The training of a SOM is an iterative process where weights of the neurons within the radius of the neighborhood function are updated according to their distance to the input data, where the neuron closest to the input data is called the Best Matching Unit (BMU). At the end of the training, the neurons are weighted so that they match the distribution of the input data.
In astronomy, SOMs have been used from classifications of stars/galaxies \cite{maehoenen_automated_1995} to measurements of the physical properties of galaxies as in \cite{hemmati_bringing_2019}, to photometric redshift measurement improvements for weak lensing cosmology as in  \cite{masters_mapping_2015,som_photoz1,som_photoz2}. 

In this work, we use SOMPY\footnote{\url{https://github.com/sevamoo/SOMPY}} (\cite{moosavi2014sompy} - a python library for SOM) with a rectangular grid. 
Each cell or neuron on the grid has a weight vector with the size of the input data set dimensions. We assign the initial weights through PCA and tune them in each epoch of the training phase by  comparing them with the training data using an Euclidean metric. 
Our choice of distance metrics is currently limited by metrics available in the SOMPY package. Alternative metrics taking into account the covariance on data vector measurements will be explored in future work.
The training is considered done when the weights of the cells represent the training data space well. 
In our case, we use the default limit radius of the neighborhood function in SOMPY as a stopping point for the training.  

The choice of grid size is not well-defined and depends mostly on the size and variations of the training sample.
A grid axis ratio of 1:1 has been shown in \cite{davidzon_horizon-agn_2019} to slightly outperform the non-square choices, we hence choose a $6 \times 6$ cells grid here, which satisfies the needs of this study (we tested that a larger grid led to a large fraction of empty cells and a smaller grid didn't allow to show the differences between theories).

\textit{Training set --} 
We use theoretical predictions of the cosmic shear real-space correlation functions $\xi_{\pm}^{ij}(\theta)$ at angular separation $\theta$ in redshift bins $i$ and $j$.
The SOM is trained on theoretical predictions of $\xi_{\pm}^{ij}(\theta)$ in various modified gravity (MG) theories, for different values of their parameters.
We model the correlation functions $\xi_{\pm}^{ij}(\theta)$ as: \begin{equation}
    \xi_{\pm}^{ij}(\theta) = \int \frac{d\ell \ell}{2 \pi} J_{0/4}(\ell \theta) C_{\ell}^{\kappa_i \kappa_j},
\end{equation} where $J_{0/4}$ are the Bessel functions of order 0 and 4, and $C_{\ell}^{\kappa_i \kappa_j}$ is the convergence $\kappa$ power spectrum between redshift bins $i$ and $j$: \begin{equation}
   C_{\ell}^{\kappa_i \kappa_j} = \int d\chi \frac{q_{\kappa}^i q_{\kappa}^j}{\chi^2} P(k = \frac{\ell + 0.5}{\chi},z(\chi)),
\end{equation}
where $\chi$ is comoving distance, $P(k,z)$ the matter power spectrum and $q_{\kappa}^i$ is the radial weight function: $q_{\kappa}^i = \frac{3H_0^2 \Omega_m \chi}{2a(\chi)} \int d\chi' \frac{\chi ' - \chi}{\chi} n^i(z(\chi')) \frac{dz}{d \chi'}$.

\begin{figure}[h]
\centering
    \includegraphics[width=0.6\columnwidth]{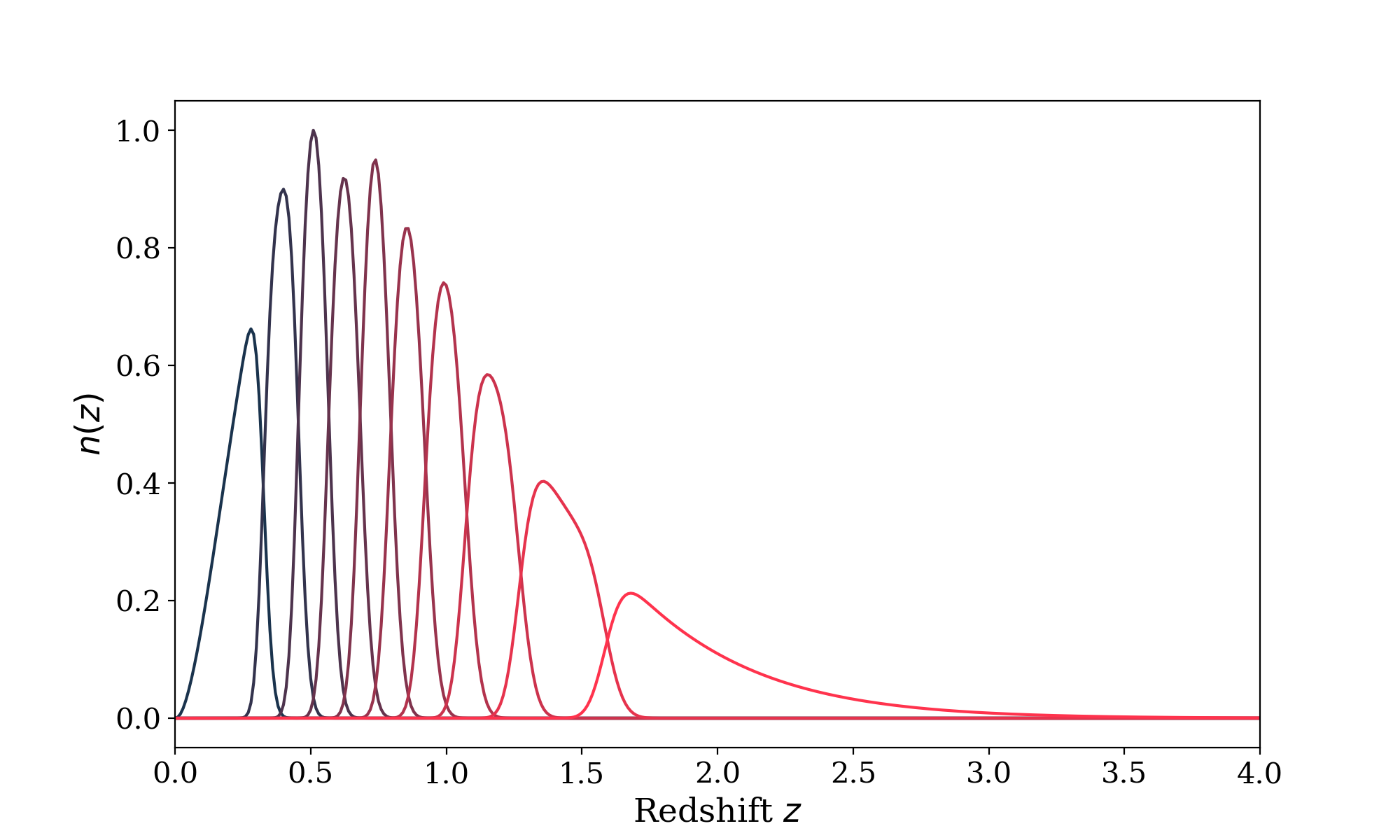}

    \caption{Redshift distribution of galaxies $n(z)$ used to compute the cosmic shear correlation functions {$\xi_{\pm}(\theta)$} forming the training set.}
    \label{fig:nz_lsst}
\end{figure}

To determine the redshift distribution $n(z)$ entering the radial weight function, we consider galaxies distributed in 10 redshift bins, similar to what is expected for LSST Year 10 \cite{lsst_forecast}. 
The forecasted binned redshift distribution is shown on figure \ref{fig:nz_lsst}, spanning from redshift 0 to 4, following $z^{\alpha} \exp{-(z/z_0)^{\beta}}$ with $\alpha = 2$, $\beta = 1$ and $z_0 = 0.3$, convolved by a Gaussian photometric error of variance $\sigma$ = 0.02 at $z = 0$.
We use the cosmological code CosmoSIS\footnote{\url{https://bitbucket.org/joezuntz/cosmosis/wiki/Home}} \cite{zuntz_cosmosis_2015} to compute the theoretical predictions of cosmic shear, where the impact of modified gravity on the matter power spectrum $P(k,z)$ is computed using the latest version of MGCAMB\footnote{\url{https://github.com/sfu-cosmo/MGCAMB}} \cite{zucca_mgcamb_2019,hojjati_testing_2011,zhao_searching_2009}, implemented in CosmoSIS.
For the purpose of this first study we assume a linear matter power spectrum.
Indeed, while various approaches have been proposed to model non-linearities in modified gravity theories, an alternative is to remove measurements of cosmic shear on non-linear scales for the cosmological analysis (as done in \cite{des_collaboration_dark_2018} for instance).
In a similar way, we use the cosmic shear correlation functions in a range of $\theta$ roughly similar to the conservative scale cuts used for the constraints on deviations to GR in \cite{des_collaboration_dark_2018}\footnote{Such scale cuts depend on the analysis' specifications (such as the number of redshift bins or binning in $\theta$). We choose the scale cuts from \cite{des_collaboration_dark_2018} for simplicity, the impact of non-linearities will be described in future work.}. 
We therefore use predictions of $\xi_+$ between 30 and 300 arcminutes and of $\xi_-$ between 250 and 300 arcminutes for all redshift combinations.
All the cosmic shear correlations are then turned into a one-dimensional vector of 45,375 elements (corresponding to 765 values of $\theta$ for $\xi+$, 60 for $\xi_-$ in 55 redshift bin correlations.)

We consider modified gravity theories that are readily available in MGCAMB and have an impact on cosmic probes, following \cite{Hojjati_2016}, and compute $\xi_{\pm}$ for 100 different values of the corresponding MG parameters. 
These models reproduce a cosmological constant in the background evolution of the recent Universe.
These values are linearly spaced within a range that we choose to not be informed by astrophysical constraints as cosmological analyses often use such large priors.
We use: \\
- The Hu-Sawicki model of $f(R)$ gravity \cite{Hu_2007} which can describe a cosmic acceleration while satisfying solar system tests of gravity.
In this model, the GR action is modified by a function of the Ricci scalar $R$.
This model is described by the Compton wavelength parameter $B_0$ \cite{fr_b0,dossett_constraining_2014,planck_2015_mg}: we vary
$\log_{10}{B_0}$ linearly between -10 and -2. We focus on this parametrisation as one of the tightest cosmological constraints on $f(\mathrm{R})$ is made on $B_0$ in \cite{planck_2015_mg}. 
An alternative $f(\mathrm{R})$ parametrisation uses \fro, which corresponds to the value of $\mathrm{d}f(R)/\mathrm{d}R$ today and $n$, the index of the power law $f(R)$ follows. 
We produced a training set for this parametrisation which is available on request, although we focus on the $B_0$ parametrisation here. \\
- The dilaton model \cite{dilaton_screen,dilaton_cosmology} parametrized by the value of the force range and the coupling to matter today, respectively $\xi_0$ and $\beta_0$. We vary $\log_{10}{\xi_0}$ linearly between -6 and -3.3 (above this value our current pipeline gives a negative shear correlated functions) for $\beta_0$ set to $\Omega_{\Lambda,0}/\Omega_{m,0}$ = 2.2 as indicated in \cite{Hojjati_2016} (for the scalar field to explain dark energy).  \\
- The symmetron model \cite{symmetron_cosmology} parametrized by the force range $\xi_{\star}$, the scale factor at the time of the force activation $a_{\star}$ and the coupling to matter $\beta_{\star}$. We vary $\log_{10}(\xi_{\star})$ between -6 and -2 for $\beta_{\star}$ set to 0.5, 1 and 1.5, while we fix $a_{\star}$ for simplicity, as is done in \cite{Hojjati_2016} for instance. We choose to set $a_{\star}$ to 0.5 as it is the standard choice, but note that cosmic shear depends on the value of $a_{\star}$, a different choice would thus lead to a different SOM.\\
\begin{table}[]
    \centering
    \begin{tabular}{|c|c|c|c|}
        \hline
        \textbf{Model} & \textbf{Parameters} & \textbf{Prior range} & \textbf{Current cosmological constraints} \\
        \hline 
        \hline
        f(R) &  $B_0$  &  [$10^{-10}$,$10^{-2}$] & $B_0 < 0.86\times 10^{-4}$ \\
        \hline
        Dilaton & $\xi_0,\beta_0 $ & [$10^{-6}$,$5\times10^{-4}$], 2.2  & $\xi_0 <$ 3$\times 10^{-3}$ \\
        \hline
        Symmetron & $\xi_{*},\beta_*,a_* $ & [$10^{-6}$,$10^{-2}$], [0.5,1.5], 0.5& $\xi_{*} <$ 1.8$\times 10^{-3}$ \\
        \hline
    \end{tabular}
    \caption{Parameters varied, their value range and current upper bounds (at 95$\%$ confidence level) for the modified gravity models used to compute the SOM training data set (in the dilaton case, $\beta_0$ is fixed to $\Omega_{\Lambda,0}/\Omega_{m,0}$). \vspace{0.1cm}}
    \label{tab:mgparams}
\end{table}
The parameters' range used to produce our training set mostly contain values of modified gravity parameters still allowed by cosmological measurements.
Firstly, \cite{planck_2015_mg} placed an upper bound on the $B_0$ parameter of $f(R)$ gravity.
In this case the 95\% confidence level upper limit is $B_0 < 0.86\times 10^{-4}$ when combining temperature and polarisation CMB measurements with cosmic shear, baryonic acoustic oscillation (BAO) and redshift space distortion measurements.
We note that cosmic shear measurements were used to constrain \fro and $n$ in \cite{Vazsonyi_2021} using HSC data and \cite{troster_kids-1000_2020} using KiDS-1000 data, and in both cosmological analyses, the results were prior dominated.
Cosmic shear data were also used in \cite{cfhtlens_fr} leading to rejection of $f(R)$ gravity with \fro $= 10^{-4}$, however at fixed $n$. 
Furthermore, using screening properties, tests of gravity on astrophysical scales have also been developed, for instance in \cite{Jain_2013} using local distance indicators and \cite{Desmond_2020} using galaxy morphology. 
These analyses obtained much more stringent upper bounds on \fro, ruling out models with \fro above $\sim 10^{-7}$.
Cosmological tests are however still relevant, at least to reproduce such constraints with future large scale structures data.
Secondly, \cite{Hojjati_2016} used CMB measurements in addition to BAO, cosmic shear and matter power spectrum available at the time leading to 95\% confidence limit upper bound of $3\times 10^{-3}$ on $\xi_0$ for the dilaton and of $1.8 \times 10^{-3}$ for $\xi_{\star}$ in the symmetron model, when varying the neutrino mass.
Table \ref{tab:mgparams} summarizes the varied parameters, their corresponding ranges and current upper bounds (at 95\% confidence level) for each of the considered MG theories.
We therefore compute shear correlation functions for a total of 500 models.

\section{Results and discussion}   

\begin{figure}
    \centering
    \includegraphics[width=0.4\textwidth]{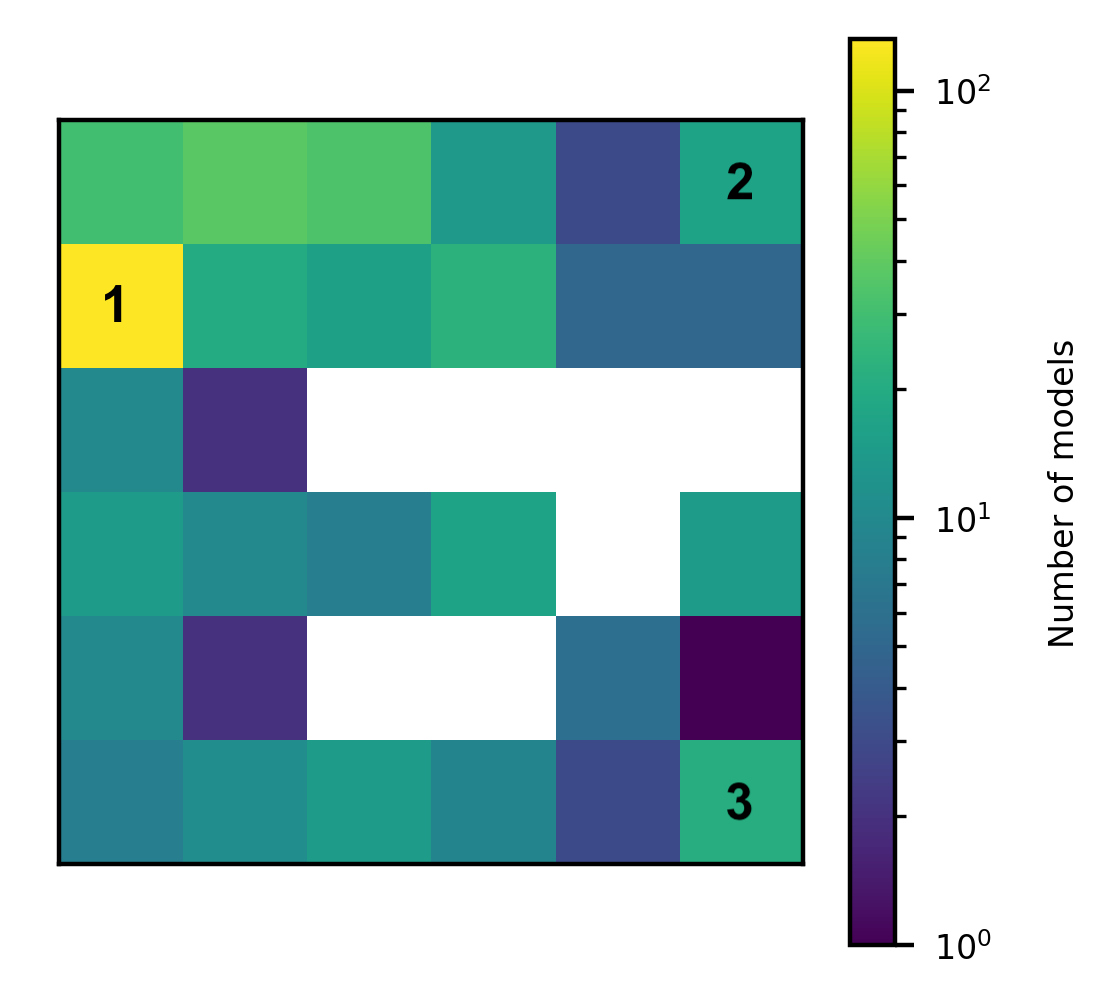}
    \caption{Self-Organizing Map trained on shear correlation functions $\xi_{\pm} (\theta)$ in $f(R)$ gravity and the dilaton and symmetron models, where cells are colored by the number of models per BMU of the SOM. Shear correlation functions $\xi_+^{22}(\theta)$ corresponding to cells numbered from 1 to 3 are represented on Fig. \ref{fig:som_mg_xip}.}
    \label{fig:som_mg_density}
\end{figure}

\textit{SOM of modified gravity theories --} After training the $6 \times 6$ square grid SOM on the training set described above, we map the training set on the SOM. Fig. \ref{fig:som_mg_density} shows the resulting two dimensional SOM grid with cells colored by the final number of models per neuron. 
In this figure and the following ones, cells in white correspond to cells with no models. 
The cell corresponding to the highest number of models lie in the left top corner (numbered 1 in Fig. \ref{fig:som_mg_density} with a total of 132 models in this cell), which corresponds to General Relativity or small deviations to GR. This indicates that shear correlation functions for small values of the modified gravity parameters ($B_0$, $\xi_0$ and $\xi_{\star}$) are indistinguishable from GR.

\begin{figure}[h]
    \centering
    \includegraphics[width=0.7\columnwidth]{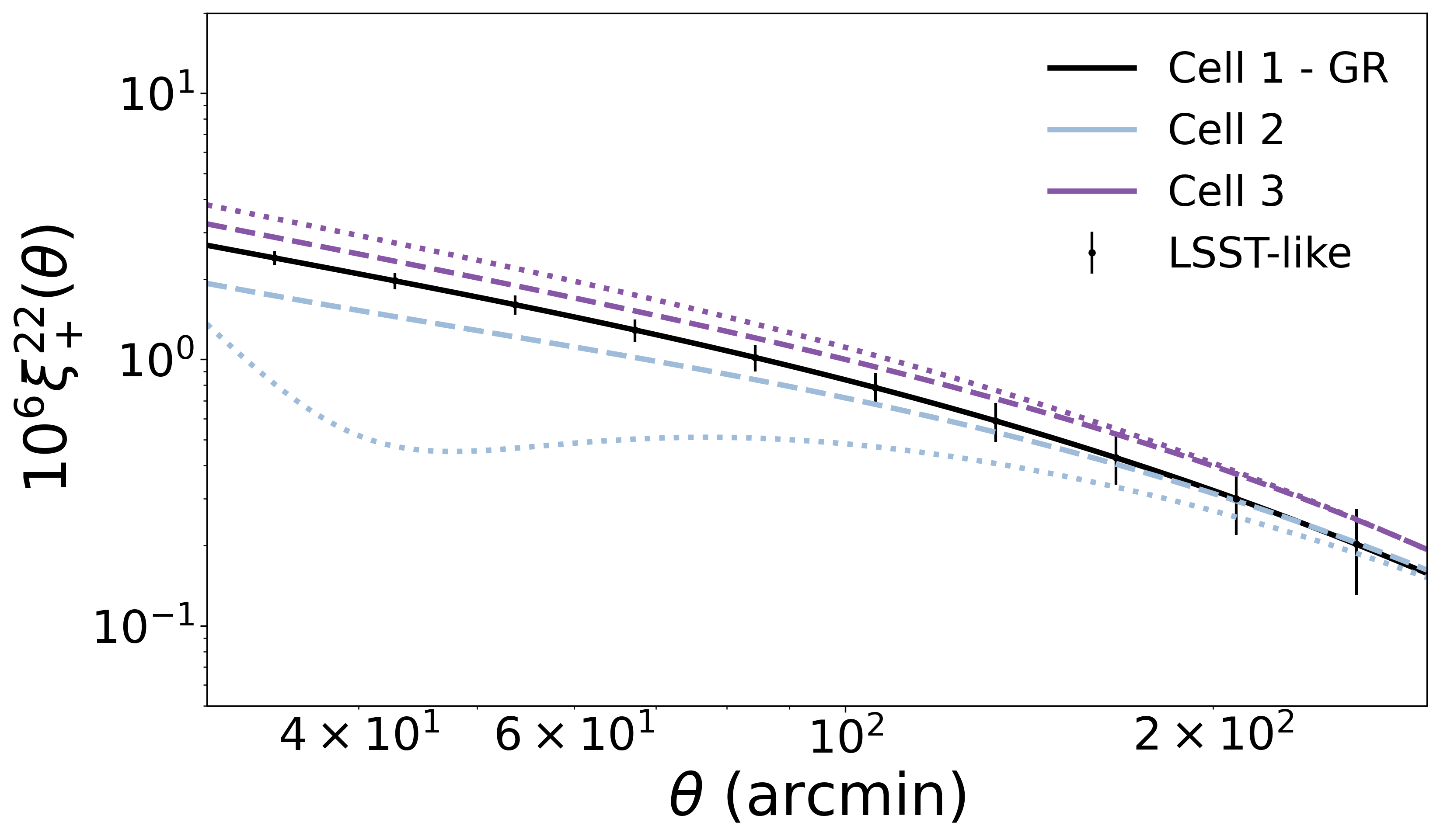}
    \caption{Shear auto-correlations $\xi^{22}_+(\theta)$ in the second redshift bin associated to the BMUs numbered from 1 to 3 on Fig. \ref{fig:som_mg_density}, respectively in black, blue and purple. The dashed and dotted lines correspond to different models in the same BMU corresponding to the closest (in dashed lines) and most distance (in dotted lines) model to the cosmic shear prediction in GR within one cell. For purpose of illustration, the black error bars show the expected error bars on $\xi_+^{22}$ for a survey like LSST.}

    \label{fig:som_mg_xip}
\end{figure}

Fig. \ref{fig:som_mg_xip} shows correlation functions $\xi^{22}_+(\theta)$ in redshift bin 2 for different cells (which are numbered from 1 to 3 in Fig. \ref{fig:som_mg_density}). 
The dotted and dashed lines for cells 2 and 3 respectively show the closest and furthest (according to the Euclidean distance metric) models associated to these two cells to GR. 
The black line respresents the median of models in cell 1, these models present negligible variations as the MG parameters are very small (e.g. the median of $B_0$ in cell 1 is 3.1$\times 10^{-9}$). 
We also show for purpose of illustration the expected error bars on $\xi_+^{22}(\theta)$ from a survey like LSST (10 source redshift bin with 55 galaxies $\mathrm{arcmin}^{-2}$ and a total shape noise of 0.3 over 50$\%$ of the sky), which we computed using CosmoCov \cite{cosmocov1,cosmocov2,cosmocov3}.

\begin{figure}[h]
\centering
    \includegraphics[width=0.3\columnwidth]{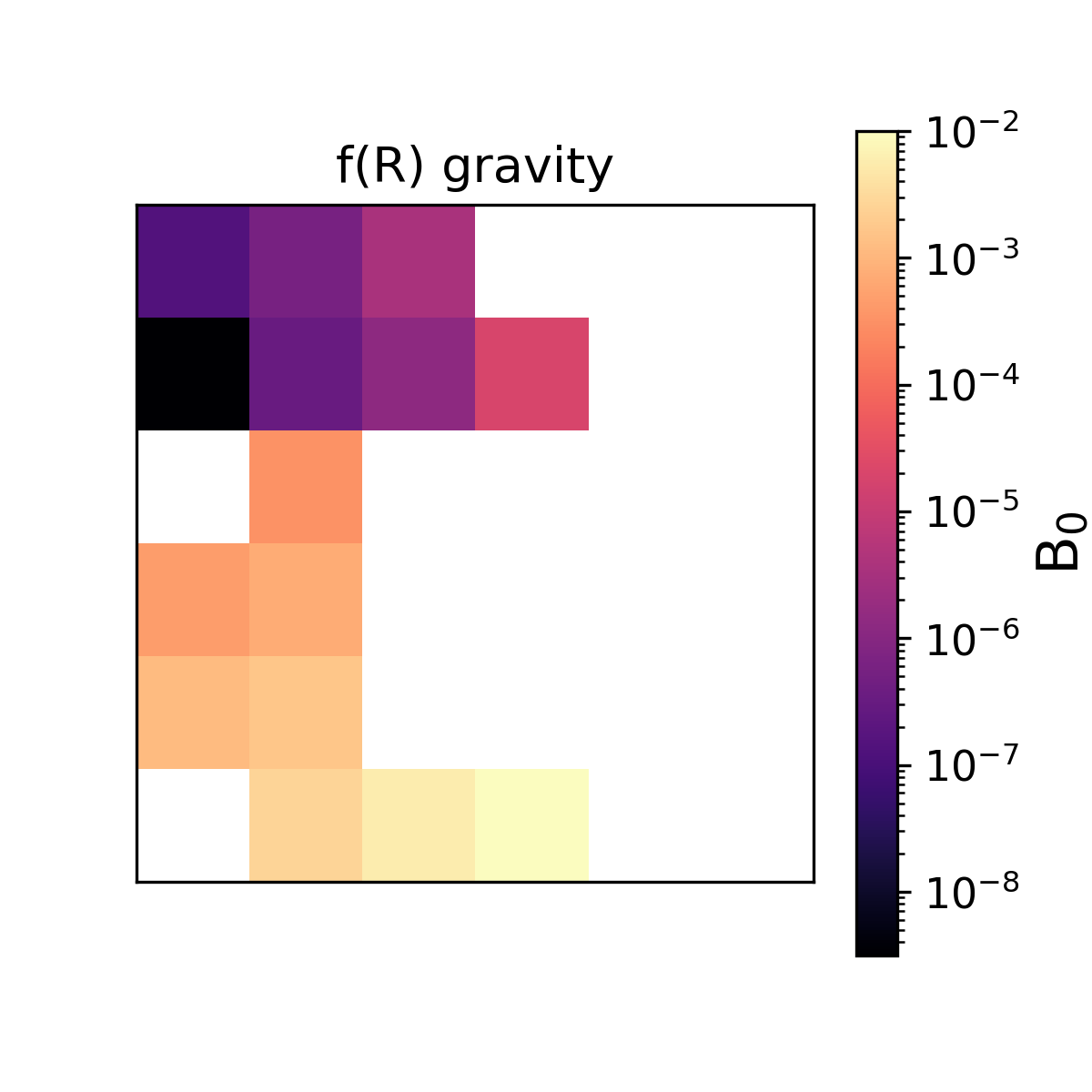}
    \includegraphics[width=0.3\columnwidth]{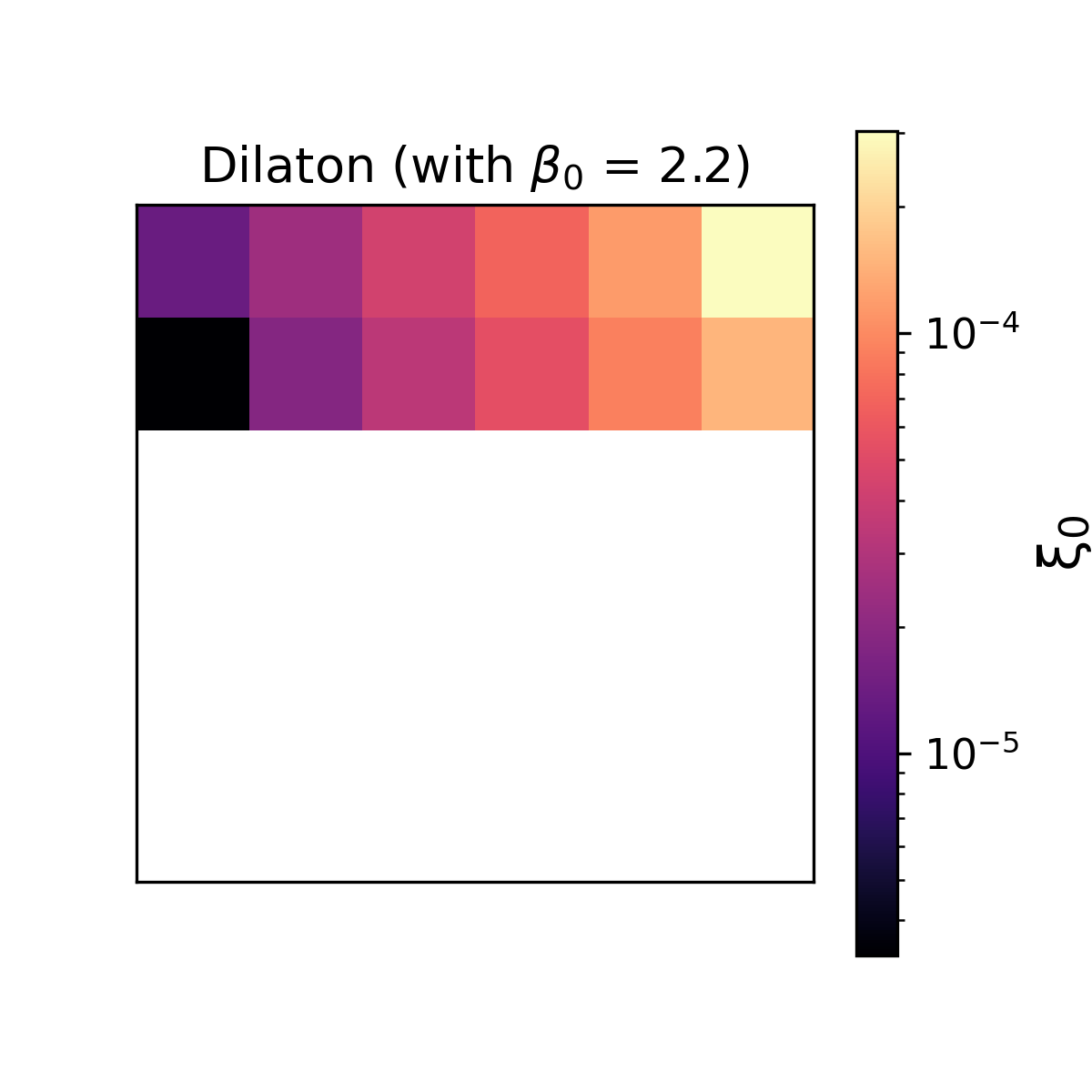}
    \includegraphics[width=0.3\columnwidth]{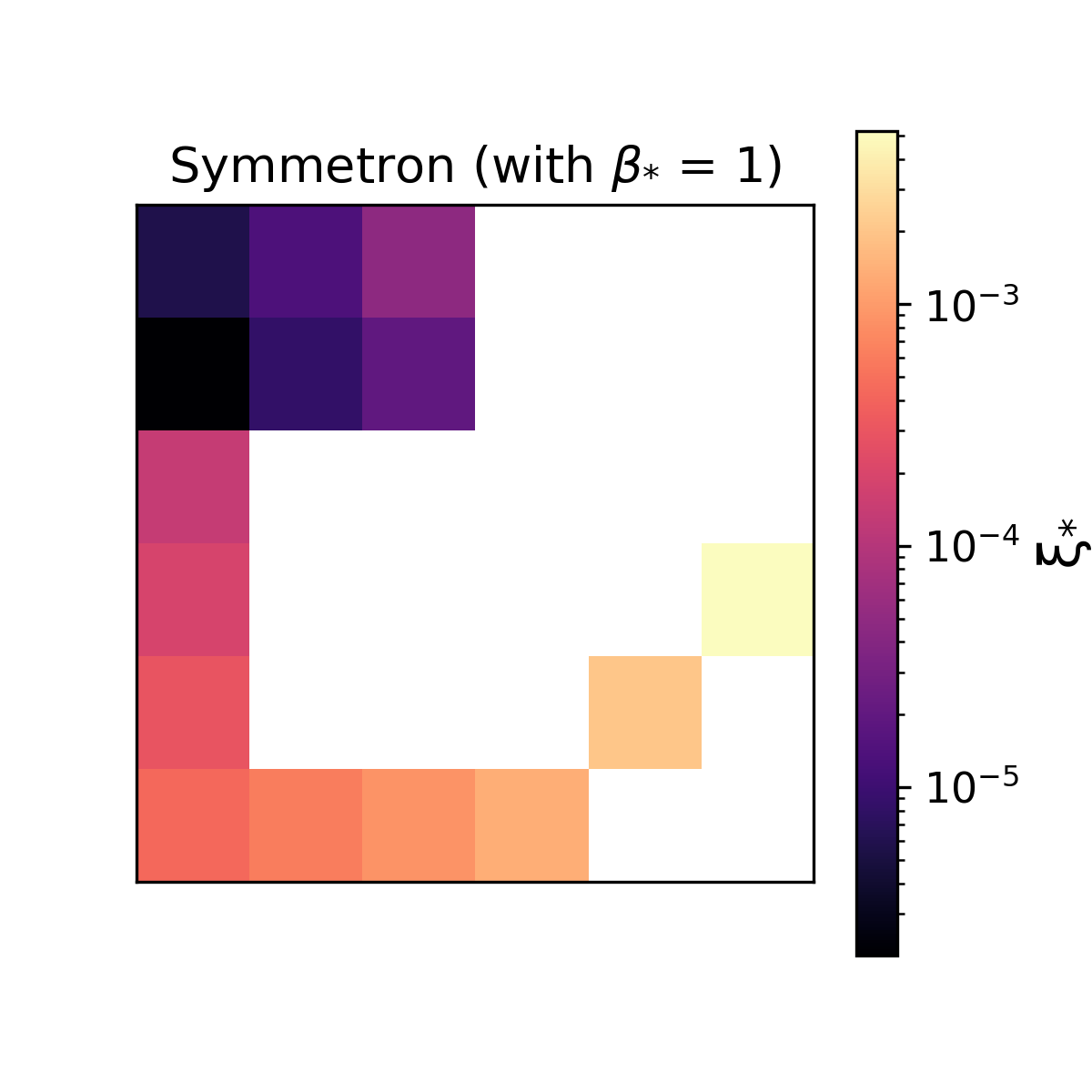}

    \caption{SOM of modified gravity from Fig. \ref{fig:som_mg_density} showing the median of the values of the parameter for each cell for $f(R)$ gravity, the dilaton (for $\beta_0$ = 2.2) and symmetron (for $\beta_{\star}$ = 1) models from left to right.
    \vspace{0.2cm}}
    \label{fig:som_mg_models}
\end{figure}

Other cells correspond to the various modified gravity theories used in the training, so we show in Fig. \ref{fig:som_mg_models} a different representation of this SOM, now highlighting only the cells corresponding to each MG theory. In this case, the cells are colored according to the median of the corresponding MG parameter in each cell.
The panels correspond to $f(R)$ gravity, the dilaton and symmetron models from left to right, only showing the distribution of $\xi_{\star}$ for $\beta_{\star}$ = 1 in the symmetron model.
The 3 theories have a large overlap in the top left part of the SOM for small values of the MG parameters. 
However, for larger values, the dilaton model correspond to cells different from those corresponding to $f(R)$ and the symmetron model. 
This means the shear vectors for the dilaton were associated to different BMUs than other models, which indicates that its signature is different enough that it should be constrained by future analysis. 
\begin{figure}
    \setlength{\lineskip}{-20pt}
    \centering
    \begin{tabular}{c}
    \hspace{1.8cm}\includegraphics[width=0.5\textwidth]{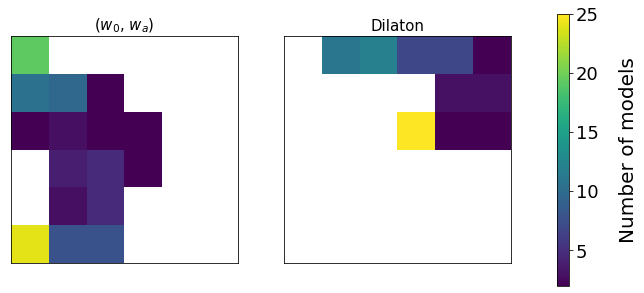} \\
    \includegraphics[width=0.4\textwidth]{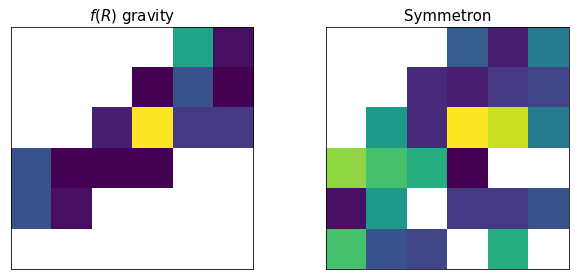}
    \end{tabular}

    \caption{Number of models in each cell for ($w_0$,$w_a$) on the top left panel and dilaton theories on the top right panel of the 6$\times$6 Self-Organizing Map grid trained on shear correlation functions $\xi_{\pm} (\theta)$ computed in ($w_0$,$w_a$) model and $f(R)$ gravity, dilaton and symmetron (for $\beta_{\star} = $ 0.5, 1 and 1.5) models. The bottom panel highlights the $f(R)$ gravity model on the left and the symmetron model on the right panel.
    \vspace{0.2cm}}
    \label{fig:som_mg_apriori}
\end{figure}
\newline
$\indent$\textit{Application: does a dynamical dark energy have the same signature as MG theories --}
Current and future galaxy surveys aim at testing if the equation of state $w$ of dark energy varies with time. The equation of state in this case often follows \cite{w0wa_1,w0wa_2} \textit{i.e.} is parametrized by $w_0$ (the value of $w$ today) and $w_a$ (its dependence with time) as: $w(a) = w_0 + w_a (1 - a(t))$
where $a(t)$ is the scale factor. 
We want to know if constraining a dynamical dark energy using cosmic shear is sufficient to explain signatures from  MG theories, using the present SOM approach.
To this end, we compute the shear correlation functions for different combinations of $w_0$ and $w_a$ using the same pipeline as used in modified gravity replacing MGCamb by CAMB \cite{Lewis:1999bs,Howlett:2012mh,camb_notes}. 
We thus produce a set of shear correlation functions corresponding to 96 combinations\footnote{the pipeline failed for some combinations of $w_0+w_a >$ 0 because of reionization properties leading to convergence failure so we get 96 combinations instead of 100.} of ($w_0$,$w_a$) randomly sampled from the chosen priors: [-1.5,-0.5] for $w_0$, [-1,1] for $w_a$. 
To do so we use the apriori sampler that is readily available in CosmoSIS.
We then train a SOM on a combination of this data set and the shear in modified gravity data set described above.
The resulting SOM is shown in Fig. \ref{fig:som_mg_apriori} with the cells corresponding to dynamical dark energy models highlighted on the left panel.
We mapped the dilaton models onto the SOM in the right panel to show that a dynamical dark energy does not overlap with the dilaton model, except for parameters values close to GR (the middle cell).
We also show cells corresponding to cosmic shear in $f(R)$ gravity on the lower left panel and the symmetron model on the lower right panel.
Although the dilaton shows the smallest overlap with ($w_0$,$w_a$), $f(R)$ gravity and the symmetron model also cover cells that are not covered by a dynamical dark energy.
This indicates these MG theories would need to be constrained in addition to a dynamical dark energy in order to extract more information from cosmic shear.



\section{Conclusion}   

We presented a new approach to categorize models using a dimensionality reduction algorithm such as Self-Organizing Map, as a promising way to help guide analyses in the case of a large theory space, like in cosmology. To summarize, one can follow the following steps to apply this approach: produce the signal expected in the different models considered, train a SOM with this training data set and inspect the resulting SOM to determine models of choice.
For this last step, we recommend imaging the number of models per cell in the final SOM.
Indeed, a large number of models in a cell or a group of cells indicates models that could not be distinguished, however cells with a low number of models correspond to one or a few models having different impact on the observable and therefore models that are worth exploring further.
More generally, one can determine clusters of models on the SOM visually or through clustering algorithms such as $k$-means which is included in SOMPY, this will be especially useful for the case of a larger number of models (such as the 74 inflationary models tested in \cite{martin_2013}).
Finally, a new data set can also be mapped on a trained SOM to determine which models can explain it.

In this paper, we applied this approach to the case of a subset of modified gravity theories, categorizing these theories through their impact on cosmic shear.
We showed for instance that the signatures left on cosmic shear from $f(R)$ gravity, the dilaton and symmetron models are similar for small deviations from GR but the dilaton can be distinguished from the other models for larger parameters values, while signatures from $f(R)$ gravity are similar to the ones left by the symmetron model for $\beta_{\star} = 1$. 
Moreover we showed MG theories have a different impact on cosmic shear compared to a dynamical dark energy, indicating that MG models should indeed be explored by future surveys such as LSST or Euclid in order to further exploit cosmic shear data.
However, this first analysis has several caveats: we assume a linear matter power spectrum, consider a fixed cosmology and we use theoretical predictions of the shear correlation functions $\xi_{\pm}(\theta)$ in finely binned $\theta$. 
We will go beyond some of these assumptions in future work and develop several applications.
As such, this approach can be applied to other cases in comsology such as cosmic inflation or any case of a large theory space in physics, in order to easily identify interesting directions of exploration.

\section*{Acknowledgement}
The research was carried out at the Jet Propulsion Laboratory, California Institute of Technology, under a contract with the National Aeronautics and Space Administration (80NM0018D0004).  In particular, this work was initiated with a JPL Data Science Pilot grant under a program run by Daniel Crichton and Richard Doyle who provided early support and advice. PLT acknowledges support for this work from a NASA Postdoctoral Program Fellowship. BM was supported by the JPL/Caltech Summer Undergraduate Research Fellowship Program.
AF would like to thank Benjamin Giblin for indicating an issue in the implementation of MGCAMB in CosmoSIS to compute $f(R)$ gravity cosmic shear predictions and Joe Zuntz for his help to fix this issue; Xiao Fang for help with CosmoCov; members of the JPL Dark Sector group and of the cosmology group at Caltech for their support and inputs; and early audiences of this work for useful discussions.

\bibliographystyle{apsrev4-2}
\bibliography{main}

\begin{thebibliography}{58}%
\makeatletter
\providecommand \@ifxundefined [1]{%
 \@ifx{#1\undefined}
}%
\providecommand \@ifnum [1]{%
 \ifnum #1\expandafter \@firstoftwo
 \else \expandafter \@secondoftwo
 \fi
}%
\providecommand \@ifx [1]{%
 \ifx #1\expandafter \@firstoftwo
 \else \expandafter \@secondoftwo
 \fi
}%
\providecommand \natexlab [1]{#1}%
\providecommand \enquote  [1]{``#1''}%
\providecommand \bibnamefont  [1]{#1}%
\providecommand \bibfnamefont [1]{#1}%
\providecommand \citenamefont [1]{#1}%
\providecommand \href@noop [0]{\@secondoftwo}%
\providecommand \href [0]{\begingroup \@sanitize@url \@href}%
\providecommand \@href[1]{\@@startlink{#1}\@@href}%
\providecommand \@@href[1]{\endgroup#1\@@endlink}%
\providecommand \@sanitize@url [0]{\catcode `\\12\catcode `\$12\catcode
  `\&12\catcode `\#12\catcode `\^12\catcode `\_12\catcode `\%12\relax}%
\providecommand \@@startlink[1]{}%
\providecommand \@@endlink[0]{}%
\providecommand \url  [0]{\begingroup\@sanitize@url \@url }%
\providecommand \@url [1]{\endgroup\@href {#1}{\urlprefix }}%
\providecommand \urlprefix  [0]{URL }%
\providecommand \Eprint [0]{\href }%
\providecommand \doibase [0]{https://doi.org/}%
\providecommand \selectlanguage [0]{\@gobble}%
\providecommand \bibinfo  [0]{\@secondoftwo}%
\providecommand \bibfield  [0]{\@secondoftwo}%
\providecommand \translation [1]{[#1]}%
\providecommand \BibitemOpen [0]{}%
\providecommand \bibitemStop [0]{}%
\providecommand \bibitemNoStop [0]{.\EOS\space}%
\providecommand \EOS [0]{\spacefactor3000\relax}%
\providecommand \BibitemShut  [1]{\csname bibitem#1\endcsname}%
\let\auto@bib@innerbib\@empty
\bibitem [{\citenamefont {Martin}\ \emph {et~al.}(2013)\citenamefont {Martin},
  \citenamefont {Ringeval},\ and\ \citenamefont {Vennin}}]{martin_2013}%
  \BibitemOpen
  \bibfield  {author} {\bibinfo {author} {\bibfnamefont {J.}~\bibnamefont
  {Martin}}, \bibinfo {author} {\bibfnamefont {C.}~\bibnamefont {Ringeval}},\
  and\ \bibinfo {author} {\bibfnamefont {V.}~\bibnamefont {Vennin}},\
  }\href@noop {} {\bibinfo {title} {Encyclopaedia inflationaris}} (\bibinfo
  {year} {2013}),\ \Eprint {https://arxiv.org/abs/1303.3787} {arXiv:1303.3787
  [astro-ph.CO]} \BibitemShut {NoStop}%
\bibitem [{\citenamefont {Ade}\ \emph {et~al.}(2014)\citenamefont {Ade},
  \citenamefont {Aghanim}, \citenamefont {Alves}, \citenamefont
  {Armitage-Caplan}, \citenamefont {Arnaud}, \citenamefont {Ashdown},
  \citenamefont {Atrio-Barandela}, \citenamefont {Aumont}, \citenamefont
  {Aussel},\ and\ \citenamefont {et~al.}}]{planck_2013}%
  \BibitemOpen
  \bibfield  {author} {\bibinfo {author} {\bibfnamefont {P.~A.~R.}\
  \bibnamefont {Ade}}, \bibinfo {author} {\bibfnamefont {N.}~\bibnamefont
  {Aghanim}}, \bibinfo {author} {\bibfnamefont {M.~I.~R.}\ \bibnamefont
  {Alves}}, \bibinfo {author} {\bibfnamefont {C.}~\bibnamefont
  {Armitage-Caplan}}, \bibinfo {author} {\bibfnamefont {M.}~\bibnamefont
  {Arnaud}}, \bibinfo {author} {\bibfnamefont {M.}~\bibnamefont {Ashdown}},
  \bibinfo {author} {\bibfnamefont {F.}~\bibnamefont {Atrio-Barandela}},
  \bibinfo {author} {\bibfnamefont {J.}~\bibnamefont {Aumont}}, \bibinfo
  {author} {\bibfnamefont {H.}~\bibnamefont {Aussel}},\ and\ \bibinfo {author}
  {\bibnamefont {et~al.}},\ }\href
  {https://doi.org/10.1051/0004-6361/201321529} {\bibfield  {journal} {\bibinfo
   {journal} {Astronomy \& Astrophysics}\ }\textbf {\bibinfo {volume} {571}},\
  \bibinfo {pages} {A1} (\bibinfo {year} {2014})}\BibitemShut {NoStop}%
\bibitem [{\citenamefont {Madhusudhan}(2019)}]{Madhusudhan_2019}%
  \BibitemOpen
  \bibfield  {author} {\bibinfo {author} {\bibfnamefont {N.}~\bibnamefont
  {Madhusudhan}},\ }\href {https://doi.org/10.1146/annurev-astro-081817-051846}
  {\bibfield  {journal} {\bibinfo  {journal} {Annual Review of Astronomy and
  Astrophysics}\ }\textbf {\bibinfo {volume} {57}},\ \bibinfo {pages} {617}
  (\bibinfo {year} {2019})}\BibitemShut {NoStop}%
\bibitem [{\citenamefont {Biller}\ and\ \citenamefont
  {Bonnefoy}(2018)}]{Biller_2018}%
  \BibitemOpen
  \bibfield  {author} {\bibinfo {author} {\bibfnamefont {B.~A.}\ \bibnamefont
  {Biller}}\ and\ \bibinfo {author} {\bibfnamefont {M.}~\bibnamefont
  {Bonnefoy}},\ }\href {https://doi.org/10.1007/978-3-319-55333-7_101}
  {\bibfield  {journal} {\bibinfo  {journal} {Handbook of Exoplanets}\ ,\
  \bibinfo {pages} {2107–2135}} (\bibinfo {year} {2018})}\BibitemShut
  {NoStop}%
\bibitem [{\citenamefont {Mancarella}\ \emph {et~al.}(2022)\citenamefont
  {Mancarella}, \citenamefont {Kennedy}, \citenamefont {Bose},\ and\
  \citenamefont {Lombriser}}]{bnn}%
  \BibitemOpen
  \bibfield  {author} {\bibinfo {author} {\bibfnamefont {M.}~\bibnamefont
  {Mancarella}}, \bibinfo {author} {\bibfnamefont {J.}~\bibnamefont {Kennedy}},
  \bibinfo {author} {\bibfnamefont {B.}~\bibnamefont {Bose}},\ and\ \bibinfo
  {author} {\bibfnamefont {L.}~\bibnamefont {Lombriser}},\ }\href
  {https://doi.org/10.1103/PhysRevD.105.023531} {\bibfield  {journal} {\bibinfo
   {journal} {Phys. Rev. D}\ }\textbf {\bibinfo {volume} {105}},\ \bibinfo
  {pages} {023531} (\bibinfo {year} {2022})}\BibitemShut {NoStop}%
\bibitem [{\citenamefont {Song}(2005)}]{song_looking_2005}%
  \BibitemOpen
  \bibfield  {author} {\bibinfo {author} {\bibfnamefont {Y.-S.}\ \bibnamefont
  {Song}},\ }\href {https://doi.org/10.1103/PhysRevD.71.024026} {\bibfield
  {journal} {\bibinfo  {journal} {Phys. Rev. D}\ }\textbf {\bibinfo {volume}
  {71}},\ \bibinfo {pages} {024026} (\bibinfo {year} {2005})}\BibitemShut
  {NoStop}%
\bibitem [{\citenamefont {Ishak}\ \emph {et~al.}(2006)\citenamefont {Ishak},
  \citenamefont {Upadhye},\ and\ \citenamefont {Spergel}}]{ishak_2005}%
  \BibitemOpen
  \bibfield  {author} {\bibinfo {author} {\bibfnamefont {M.}~\bibnamefont
  {Ishak}}, \bibinfo {author} {\bibfnamefont {A.}~\bibnamefont {Upadhye}},\
  and\ \bibinfo {author} {\bibfnamefont {D.~N.}\ \bibnamefont {Spergel}},\
  }\href {https://doi.org/10.1103/PhysRevD.74.043513} {\bibfield  {journal}
  {\bibinfo  {journal} {Phys. Rev. D}\ }\textbf {\bibinfo {volume} {74}},\
  \bibinfo {pages} {043513} (\bibinfo {year} {2006})}\BibitemShut {NoStop}%
\bibitem [{\citenamefont {Song}(2006)}]{song_cosmological_2006}%
  \BibitemOpen
  \bibfield  {author} {\bibinfo {author} {\bibnamefont {Song}},\ }\href
  {http://arxiv.org/abs/astro-ph/0602598} {\bibfield  {journal} {\bibinfo
  {journal} {arXiv:astro-ph/0602598}\ } (\bibinfo {year} {2006})}\BibitemShut
  {NoStop}%
\bibitem [{\citenamefont {Tsujikawa}\ and\ \citenamefont
  {Tatekawa}(2008)}]{tsujikawa_effect_2008}%
  \BibitemOpen
  \bibfield  {author} {\bibinfo {author} {\bibfnamefont {S.}~\bibnamefont
  {Tsujikawa}}\ and\ \bibinfo {author} {\bibfnamefont {T.}~\bibnamefont
  {Tatekawa}},\ }\href {https://doi.org/10.1016/j.physletb.2008.06.052}
  {\bibfield  {journal} {\bibinfo  {journal} {Physics Letters B}\ }\textbf
  {\bibinfo {volume} {665}},\ \bibinfo {pages} {325} (\bibinfo {year}
  {2008})},\ \bibinfo {note} {arXiv: 0804.4343}\BibitemShut {NoStop}%
\bibitem [{\citenamefont {Jain}\ and\ \citenamefont
  {Zhang}(2008)}]{jain_observational_2008}%
  \BibitemOpen
  \bibfield  {author} {\bibinfo {author} {\bibfnamefont {B.}~\bibnamefont
  {Jain}}\ and\ \bibinfo {author} {\bibfnamefont {P.}~\bibnamefont {Zhang}},\
  }\href {https://doi.org/10.1103/PhysRevD.78.063503} {\bibfield  {journal}
  {\bibinfo  {journal} {Physical Review D}\ }\textbf {\bibinfo {volume} {78}},\
  \bibinfo {pages} {063503} (\bibinfo {year} {2008})},\ \bibinfo {note} {arXiv:
  0709.2375}\BibitemShut {NoStop}%
\bibitem [{\citenamefont {Harnois-Déraps}\ \emph {et~al.}(2015)\citenamefont
  {Harnois-Déraps}, \citenamefont {Munshi}, \citenamefont {Valageas},
  \citenamefont {van Waerbeke}, \citenamefont {Brax}, \citenamefont {Coles},\
  and\ \citenamefont {Rizzo}}]{cfhtlens_fr}%
  \BibitemOpen
  \bibfield  {author} {\bibinfo {author} {\bibfnamefont {J.}~\bibnamefont
  {Harnois-Déraps}}, \bibinfo {author} {\bibfnamefont {D.}~\bibnamefont
  {Munshi}}, \bibinfo {author} {\bibfnamefont {P.}~\bibnamefont {Valageas}},
  \bibinfo {author} {\bibfnamefont {L.}~\bibnamefont {van Waerbeke}}, \bibinfo
  {author} {\bibfnamefont {P.}~\bibnamefont {Brax}}, \bibinfo {author}
  {\bibfnamefont {P.}~\bibnamefont {Coles}},\ and\ \bibinfo {author}
  {\bibfnamefont {L.}~\bibnamefont {Rizzo}},\ }\href
  {https://doi.org/10.1093/mnras/stv2120} {\bibfield  {journal} {\bibinfo
  {journal} {Monthly Notices of the Royal Astronomical Society}\ }\textbf
  {\bibinfo {volume} {454}},\ \bibinfo {pages} {2722} (\bibinfo {year}
  {2015})}\BibitemShut {NoStop}%
\bibitem [{\citenamefont {Simpson}\ \emph {et~al.}(2013)\citenamefont
  {Simpson}, \citenamefont {Heymans}, \citenamefont {Parkinson}, \citenamefont
  {Blake}, \citenamefont {Kilbinger}, \citenamefont {Benjamin}, \citenamefont
  {Erben}, \citenamefont {Hildebrandt}, \citenamefont {Hoekstra}, \citenamefont
  {Kitching} \emph {et~al.}}]{simpson_cfhtlens}%
  \BibitemOpen
  \bibfield  {author} {\bibinfo {author} {\bibfnamefont {F.}~\bibnamefont
  {Simpson}}, \bibinfo {author} {\bibfnamefont {C.}~\bibnamefont {Heymans}},
  \bibinfo {author} {\bibfnamefont {D.}~\bibnamefont {Parkinson}}, \bibinfo
  {author} {\bibfnamefont {C.}~\bibnamefont {Blake}}, \bibinfo {author}
  {\bibfnamefont {M.}~\bibnamefont {Kilbinger}}, \bibinfo {author}
  {\bibfnamefont {J.}~\bibnamefont {Benjamin}}, \bibinfo {author}
  {\bibfnamefont {T.}~\bibnamefont {Erben}}, \bibinfo {author} {\bibfnamefont
  {H.}~\bibnamefont {Hildebrandt}}, \bibinfo {author} {\bibfnamefont
  {H.}~\bibnamefont {Hoekstra}}, \bibinfo {author} {\bibfnamefont {T.~D.}\
  \bibnamefont {Kitching}}, \emph {et~al.},\ }\href
  {https://doi.org/10.1093/mnras/sts493} {\bibfield  {journal} {\bibinfo
  {journal} {Monthly Notices of the Royal Astronomical Society}\ }\textbf
  {\bibinfo {volume} {429}},\ \bibinfo {pages} {2249} (\bibinfo {year}
  {2013})}\BibitemShut {NoStop}%
\bibitem [{\citenamefont {Dossett}\ \emph {et~al.}(2015)\citenamefont
  {Dossett}, \citenamefont {Ishak}, \citenamefont {Parkinson},\ and\
  \citenamefont {Davis}}]{Dossett_2015}%
  \BibitemOpen
  \bibfield  {author} {\bibinfo {author} {\bibfnamefont {J.~N.}\ \bibnamefont
  {Dossett}}, \bibinfo {author} {\bibfnamefont {M.}~\bibnamefont {Ishak}},
  \bibinfo {author} {\bibfnamefont {D.}~\bibnamefont {Parkinson}},\ and\
  \bibinfo {author} {\bibfnamefont {T.~M.}\ \bibnamefont {Davis}},\ }\href
  {https://doi.org/10.1103/PhysRevD.92.023003} {\bibfield  {journal} {\bibinfo
  {journal} {Phys. Rev. D}\ }\textbf {\bibinfo {volume} {92}},\ \bibinfo
  {pages} {023003} (\bibinfo {year} {2015})}\BibitemShut {NoStop}%
\bibitem [{\citenamefont {Fert\'e}\ \emph {et~al.}(2019)\citenamefont
  {Fert\'e}, \citenamefont {Kirk}, \citenamefont {Liddle},\ and\ \citenamefont
  {Zuntz}}]{ferte_testing_2017}%
  \BibitemOpen
  \bibfield  {author} {\bibinfo {author} {\bibfnamefont {A.}~\bibnamefont
  {Fert\'e}}, \bibinfo {author} {\bibfnamefont {D.}~\bibnamefont {Kirk}},
  \bibinfo {author} {\bibfnamefont {A.~R.}\ \bibnamefont {Liddle}},\ and\
  \bibinfo {author} {\bibfnamefont {J.}~\bibnamefont {Zuntz}},\ }\href
  {https://doi.org/10.1103/PhysRevD.99.083512} {\bibfield  {journal} {\bibinfo
  {journal} {Phys. Rev. D}\ }\textbf {\bibinfo {volume} {99}},\ \bibinfo
  {pages} {083512} (\bibinfo {year} {2019})}\BibitemShut {NoStop}%
\bibitem [{\citenamefont {Joudaki}\ \emph {et~al.}(2017)\citenamefont
  {Joudaki}, \citenamefont {Mead}, \citenamefont {Blake}, \citenamefont {Choi},
  \citenamefont {de~Jong}, \citenamefont {Erben}, \citenamefont {Conti},
  \citenamefont {Herbonnet}, \citenamefont {Heymans}, \citenamefont
  {Hildebrandt} \emph {et~al.}}]{joudaki_kids-450_2017}%
  \BibitemOpen
  \bibfield  {author} {\bibinfo {author} {\bibfnamefont {S.}~\bibnamefont
  {Joudaki}}, \bibinfo {author} {\bibfnamefont {A.}~\bibnamefont {Mead}},
  \bibinfo {author} {\bibfnamefont {C.}~\bibnamefont {Blake}}, \bibinfo
  {author} {\bibfnamefont {A.}~\bibnamefont {Choi}}, \bibinfo {author}
  {\bibfnamefont {J.}~\bibnamefont {de~Jong}}, \bibinfo {author} {\bibfnamefont
  {T.}~\bibnamefont {Erben}}, \bibinfo {author} {\bibfnamefont {I.~F.}\
  \bibnamefont {Conti}}, \bibinfo {author} {\bibfnamefont {R.}~\bibnamefont
  {Herbonnet}}, \bibinfo {author} {\bibfnamefont {C.}~\bibnamefont {Heymans}},
  \bibinfo {author} {\bibfnamefont {H.}~\bibnamefont {Hildebrandt}}, \emph
  {et~al.},\ }\href {https://doi.org/10.1093/mnras/stx998} {\bibfield
  {journal} {\bibinfo  {journal} {Monthly Notices of the Royal Astronomical
  Society}\ }\textbf {\bibinfo {volume} {471}},\ \bibinfo {pages} {1259}
  (\bibinfo {year} {2017})}\BibitemShut {NoStop}%
\bibitem [{\citenamefont {Tröster}\ \emph {et~al.}(2021)\citenamefont
  {Tröster}, \citenamefont {Asgari}, \citenamefont {Blake}, \citenamefont
  {Cataneo}, \citenamefont {Heymans}, \citenamefont {Hildebrandt},
  \citenamefont {Joachimi}, \citenamefont {Lin}, \citenamefont {S{\'{a}
  }nchez}, \citenamefont {Wright} \emph {et~al.}}]{troster_kids-1000_2020}%
  \BibitemOpen
  \bibfield  {author} {\bibinfo {author} {\bibfnamefont {T.}~\bibnamefont
  {Tröster}}, \bibinfo {author} {\bibfnamefont {M.}~\bibnamefont {Asgari}},
  \bibinfo {author} {\bibfnamefont {C.}~\bibnamefont {Blake}}, \bibinfo
  {author} {\bibfnamefont {M.}~\bibnamefont {Cataneo}}, \bibinfo {author}
  {\bibfnamefont {C.}~\bibnamefont {Heymans}}, \bibinfo {author} {\bibfnamefont
  {H.}~\bibnamefont {Hildebrandt}}, \bibinfo {author} {\bibfnamefont
  {B.}~\bibnamefont {Joachimi}}, \bibinfo {author} {\bibfnamefont {C.-A.}\
  \bibnamefont {Lin}}, \bibinfo {author} {\bibfnamefont {A.~G.}\ \bibnamefont
  {S{\'{a} }nchez}}, \bibinfo {author} {\bibfnamefont {A.~H.}\ \bibnamefont
  {Wright}}, \emph {et~al.},\ }\href
  {https://doi.org/10.1051/0004-6361/202039805} {\bibfield  {journal} {\bibinfo
   {journal} {Astronomy \& Astrophysics}\ }\textbf {\bibinfo {volume} {649}},\
  \bibinfo {pages} {A88} (\bibinfo {year} {2021})}\BibitemShut {NoStop}%
\bibitem [{\citenamefont {Abbott}\ \emph {et~al.}(2019)\citenamefont {Abbott},
  \citenamefont {Abdalla}, \citenamefont {Avila}, \citenamefont {Banerji},
  \citenamefont {Baxter}, \citenamefont {Bechtol}, \citenamefont {Becker},
  \citenamefont {Bertin}, \citenamefont {Blazek}, \citenamefont {Bridle} \emph
  {et~al.}}]{des_collaboration_dark_2018}%
  \BibitemOpen
  \bibfield  {author} {\bibinfo {author} {\bibfnamefont {T.~M.~C.}\
  \bibnamefont {Abbott}}, \bibinfo {author} {\bibfnamefont {F.~B.}\
  \bibnamefont {Abdalla}}, \bibinfo {author} {\bibfnamefont {S.}~\bibnamefont
  {Avila}}, \bibinfo {author} {\bibfnamefont {M.}~\bibnamefont {Banerji}},
  \bibinfo {author} {\bibfnamefont {E.}~\bibnamefont {Baxter}}, \bibinfo
  {author} {\bibfnamefont {K.}~\bibnamefont {Bechtol}}, \bibinfo {author}
  {\bibfnamefont {M.~R.}\ \bibnamefont {Becker}}, \bibinfo {author}
  {\bibfnamefont {E.}~\bibnamefont {Bertin}}, \bibinfo {author} {\bibfnamefont
  {J.}~\bibnamefont {Blazek}}, \bibinfo {author} {\bibfnamefont {S.~L.}\
  \bibnamefont {Bridle}}, \emph {et~al.} (\bibinfo {collaboration} {DES
  Collaboration}),\ }\href {https://doi.org/10.1103/PhysRevD.99.123505}
  {\bibfield  {journal} {\bibinfo  {journal} {Phys. Rev. D}\ }\textbf {\bibinfo
  {volume} {99}},\ \bibinfo {pages} {123505} (\bibinfo {year}
  {2019})}\BibitemShut {NoStop}%
\bibitem [{\citenamefont {Martinelli}\ \emph {et~al.}(2011)\citenamefont
  {Martinelli}, \citenamefont {Calabrese}, \citenamefont {De~Bernardis},
  \citenamefont {Melchiorri}, \citenamefont {Pagano},\ and\ \citenamefont
  {Scaramella}}]{martinelli_constraining_2011}%
  \BibitemOpen
  \bibfield  {author} {\bibinfo {author} {\bibfnamefont {M.}~\bibnamefont
  {Martinelli}}, \bibinfo {author} {\bibfnamefont {E.}~\bibnamefont
  {Calabrese}}, \bibinfo {author} {\bibfnamefont {F.}~\bibnamefont
  {De~Bernardis}}, \bibinfo {author} {\bibfnamefont {A.}~\bibnamefont
  {Melchiorri}}, \bibinfo {author} {\bibfnamefont {L.}~\bibnamefont {Pagano}},\
  and\ \bibinfo {author} {\bibfnamefont {R.}~\bibnamefont {Scaramella}},\
  }\href {https://doi.org/10.1103/PhysRevD.83.023012} {\bibfield  {journal}
  {\bibinfo  {journal} {Physical Review D}\ }\textbf {\bibinfo {volume} {83}},\
  \bibinfo {pages} {023012} (\bibinfo {year} {2011})}\BibitemShut {NoStop}%
\bibitem [{\citenamefont {{Euclid
  Collaboration}}(2020)}]{euclid_collaboration_euclid_2020}%
  \BibitemOpen
  \bibfield  {author} {\bibinfo {author} {\bibnamefont {{Euclid
  Collaboration}}},\ }\href {https://doi.org/10.1051/0004-6361/202038071}
  {\bibfield  {journal} {\bibinfo  {journal} {Astronomy and Astrophysics}\
  }\textbf {\bibinfo {volume} {642}},\ \bibinfo {pages} {A191} (\bibinfo {year}
  {2020})}\BibitemShut {NoStop}%
\bibitem [{\citenamefont {{The LSST Dark Energy Science
  Collaboration}}(2018)}]{lsst_forecast}%
  \BibitemOpen
  \bibfield  {author} {\bibinfo {author} {\bibnamefont {{The LSST Dark Energy
  Science Collaboration}}},\ }\href@noop {} {\bibinfo {title} {The lsst dark
  energy science collaboration (desc) science requirements document}} (\bibinfo
  {year} {2018}),\ \Eprint {https://arxiv.org/abs/1809.01669} {arXiv:1809.01669
  [astro-ph.CO]} \BibitemShut {NoStop}%
\bibitem [{\citenamefont {Hojjati}\ \emph {et~al.}(2016)\citenamefont
  {Hojjati}, \citenamefont {Plahn}, \citenamefont {Zucca}, \citenamefont
  {Pogosian}, \citenamefont {Brax}, \citenamefont {Davis},\ and\ \citenamefont
  {Zhao}}]{Hojjati_2016}%
  \BibitemOpen
  \bibfield  {author} {\bibinfo {author} {\bibfnamefont {A.}~\bibnamefont
  {Hojjati}}, \bibinfo {author} {\bibfnamefont {A.}~\bibnamefont {Plahn}},
  \bibinfo {author} {\bibfnamefont {A.}~\bibnamefont {Zucca}}, \bibinfo
  {author} {\bibfnamefont {L.}~\bibnamefont {Pogosian}}, \bibinfo {author}
  {\bibfnamefont {P.}~\bibnamefont {Brax}}, \bibinfo {author} {\bibfnamefont
  {A.-C.}\ \bibnamefont {Davis}},\ and\ \bibinfo {author} {\bibfnamefont
  {G.-B.}\ \bibnamefont {Zhao}},\ }\href
  {https://doi.org/10.1103/PhysRevD.93.043531} {\bibfield  {journal} {\bibinfo
  {journal} {Phys. Rev. D}\ }\textbf {\bibinfo {volume} {93}},\ \bibinfo
  {pages} {043531} (\bibinfo {year} {2016})}\BibitemShut {NoStop}%
\bibitem [{\citenamefont {Doré}\ \emph {et~al.}(2019)\citenamefont {Doré},
  \citenamefont {Hirata}, \citenamefont {Wang}, \citenamefont {Weinberg},
  \citenamefont {Eifler}, \citenamefont {Foley}, \citenamefont {Heinrich},
  \citenamefont {Krause}, \citenamefont {Perlmutter}, \citenamefont {Pisani}
  \emph {et~al.}}]{dore_wfirst_2019}%
  \BibitemOpen
  \bibfield  {author} {\bibinfo {author} {\bibfnamefont {O.}~\bibnamefont
  {Doré}}, \bibinfo {author} {\bibfnamefont {C.}~\bibnamefont {Hirata}},
  \bibinfo {author} {\bibfnamefont {Y.}~\bibnamefont {Wang}}, \bibinfo {author}
  {\bibfnamefont {D.}~\bibnamefont {Weinberg}}, \bibinfo {author}
  {\bibfnamefont {T.}~\bibnamefont {Eifler}}, \bibinfo {author} {\bibfnamefont
  {R.~J.}\ \bibnamefont {Foley}}, \bibinfo {author} {\bibfnamefont {C.~H.}\
  \bibnamefont {Heinrich}}, \bibinfo {author} {\bibfnamefont {E.}~\bibnamefont
  {Krause}}, \bibinfo {author} {\bibfnamefont {S.}~\bibnamefont {Perlmutter}},
  \bibinfo {author} {\bibfnamefont {A.}~\bibnamefont {Pisani}}, \emph
  {et~al.},\ }\href {http://arxiv.org/abs/1904.01174} {\bibfield  {journal}
  {\bibinfo  {journal} {arXiv:1904.01174 [astro-ph]}\ } (\bibinfo {year}
  {2019})}\BibitemShut {NoStop}%
\bibitem [{\citenamefont {Eifler}\ \emph
  {et~al.}(2021{\natexlab{a}})\citenamefont {Eifler}, \citenamefont {Miyatake},
  \citenamefont {Krause}, \citenamefont {Heinrich}, \citenamefont {Miranda},
  \citenamefont {Hirata}, \citenamefont {Xu}, \citenamefont {Hemmati},
  \citenamefont {Simet}, \citenamefont {Capak}, \citenamefont {Choi},
  \citenamefont {Dor{\'{e} }}, \citenamefont {Doux}, \citenamefont {Fang},
  \citenamefont {Hounsell}, \citenamefont {Huff}, \citenamefont {Huang},
  \citenamefont {Jarvis}, \citenamefont {Kruk}, \citenamefont {Masters},
  \citenamefont {Rozo}, \citenamefont {Scolnic}, \citenamefont {Spergel},
  \citenamefont {Troxel}, \citenamefont {von~der Linden}, \citenamefont {Wang},
  \citenamefont {Weinberg}, \citenamefont {Wenzl},\ and\ \citenamefont
  {Wu}}]{eifler_cosmology_2020}%
  \BibitemOpen
  \bibfield  {author} {\bibinfo {author} {\bibfnamefont {T.}~\bibnamefont
  {Eifler}}, \bibinfo {author} {\bibfnamefont {H.}~\bibnamefont {Miyatake}},
  \bibinfo {author} {\bibfnamefont {E.}~\bibnamefont {Krause}}, \bibinfo
  {author} {\bibfnamefont {C.}~\bibnamefont {Heinrich}}, \bibinfo {author}
  {\bibfnamefont {V.}~\bibnamefont {Miranda}}, \bibinfo {author} {\bibfnamefont
  {C.}~\bibnamefont {Hirata}}, \bibinfo {author} {\bibfnamefont
  {J.}~\bibnamefont {Xu}}, \bibinfo {author} {\bibfnamefont {S.}~\bibnamefont
  {Hemmati}}, \bibinfo {author} {\bibfnamefont {M.}~\bibnamefont {Simet}},
  \bibinfo {author} {\bibfnamefont {P.}~\bibnamefont {Capak}}, \bibinfo
  {author} {\bibfnamefont {A.}~\bibnamefont {Choi}}, \bibinfo {author}
  {\bibfnamefont {O.}~\bibnamefont {Dor{\'{e} }}}, \bibinfo {author}
  {\bibfnamefont {C.}~\bibnamefont {Doux}}, \bibinfo {author} {\bibfnamefont
  {X.}~\bibnamefont {Fang}}, \bibinfo {author} {\bibfnamefont {R.}~\bibnamefont
  {Hounsell}}, \bibinfo {author} {\bibfnamefont {E.}~\bibnamefont {Huff}},
  \bibinfo {author} {\bibfnamefont {H.-J.}\ \bibnamefont {Huang}}, \bibinfo
  {author} {\bibfnamefont {M.}~\bibnamefont {Jarvis}}, \bibinfo {author}
  {\bibfnamefont {J.}~\bibnamefont {Kruk}}, \bibinfo {author} {\bibfnamefont
  {D.}~\bibnamefont {Masters}}, \bibinfo {author} {\bibfnamefont
  {E.}~\bibnamefont {Rozo}}, \bibinfo {author} {\bibfnamefont {D.}~\bibnamefont
  {Scolnic}}, \bibinfo {author} {\bibfnamefont {D.~N.}\ \bibnamefont
  {Spergel}}, \bibinfo {author} {\bibfnamefont {M.}~\bibnamefont {Troxel}},
  \bibinfo {author} {\bibfnamefont {A.}~\bibnamefont {von~der Linden}},
  \bibinfo {author} {\bibfnamefont {Y.}~\bibnamefont {Wang}}, \bibinfo {author}
  {\bibfnamefont {D.~H.}\ \bibnamefont {Weinberg}}, \bibinfo {author}
  {\bibfnamefont {L.}~\bibnamefont {Wenzl}},\ and\ \bibinfo {author}
  {\bibfnamefont {H.-Y.}\ \bibnamefont {Wu}},\ }\href
  {https://doi.org/10.1093/mnras/stab1762} {\bibfield  {journal} {\bibinfo
  {journal} {Monthly Notices of the Royal Astronomical Society}\ }\textbf
  {\bibinfo {volume} {507}},\ \bibinfo {pages} {1746} (\bibinfo {year}
  {2021}{\natexlab{a}})}\BibitemShut {NoStop}%
\bibitem [{\citenamefont {Eifler}\ \emph
  {et~al.}(2021{\natexlab{b}})\citenamefont {Eifler}, \citenamefont {Simet},
  \citenamefont {Krause}, \citenamefont {Hirata}, \citenamefont {Huang},
  \citenamefont {Fang}, \citenamefont {Miranda}, \citenamefont {Mandelbaum},
  \citenamefont {Doux}, \citenamefont {Heinrich} \emph
  {et~al.}}]{eifler_cosmology_2021}%
  \BibitemOpen
  \bibfield  {author} {\bibinfo {author} {\bibfnamefont {T.}~\bibnamefont
  {Eifler}}, \bibinfo {author} {\bibfnamefont {M.}~\bibnamefont {Simet}},
  \bibinfo {author} {\bibfnamefont {E.}~\bibnamefont {Krause}}, \bibinfo
  {author} {\bibfnamefont {C.}~\bibnamefont {Hirata}}, \bibinfo {author}
  {\bibfnamefont {H.-J.}\ \bibnamefont {Huang}}, \bibinfo {author}
  {\bibfnamefont {X.}~\bibnamefont {Fang}}, \bibinfo {author} {\bibfnamefont
  {V.}~\bibnamefont {Miranda}}, \bibinfo {author} {\bibfnamefont
  {R.}~\bibnamefont {Mandelbaum}}, \bibinfo {author} {\bibfnamefont
  {C.}~\bibnamefont {Doux}}, \bibinfo {author} {\bibfnamefont {C.}~\bibnamefont
  {Heinrich}}, \emph {et~al.},\ }\href {https://doi.org/10.1093/mnras/stab533}
  {\bibfield  {journal} {\bibinfo  {journal} {Monthly Notices of the Royal
  Astronomical Society}\ ,\ \bibinfo {pages} {stab533}} (\bibinfo {year}
  {2021}{\natexlab{b}})}\BibitemShut {NoStop}%
\bibitem [{\citenamefont {Ishak}\ \emph {et~al.}(2019)\citenamefont {Ishak},
  \citenamefont {Baker}, \citenamefont {Bull}, \citenamefont {Pedersen},
  \citenamefont {Blazek}, \citenamefont {Ferreira}, \citenamefont {Leonard},
  \citenamefont {Lin}, \citenamefont {Linder}, \citenamefont {Pardo},\ and\
  \citenamefont {Valogiannis}}]{lsst_mg}%
  \BibitemOpen
  \bibfield  {author} {\bibinfo {author} {\bibfnamefont {M.}~\bibnamefont
  {Ishak}}, \bibinfo {author} {\bibfnamefont {T.}~\bibnamefont {Baker}},
  \bibinfo {author} {\bibfnamefont {P.}~\bibnamefont {Bull}}, \bibinfo {author}
  {\bibfnamefont {E.~M.}\ \bibnamefont {Pedersen}}, \bibinfo {author}
  {\bibfnamefont {J.}~\bibnamefont {Blazek}}, \bibinfo {author} {\bibfnamefont
  {P.~G.}\ \bibnamefont {Ferreira}}, \bibinfo {author} {\bibfnamefont {C.~D.}\
  \bibnamefont {Leonard}}, \bibinfo {author} {\bibfnamefont {W.}~\bibnamefont
  {Lin}}, \bibinfo {author} {\bibfnamefont {E.}~\bibnamefont {Linder}},
  \bibinfo {author} {\bibfnamefont {K.}~\bibnamefont {Pardo}},\ and\ \bibinfo
  {author} {\bibfnamefont {G.}~\bibnamefont {Valogiannis}},\ }\href@noop {}
  {\bibinfo {title} {Modified gravity and dark energy models beyond $w(z)$cdm
  testable by lsst}} (\bibinfo {year} {2019}),\ \Eprint
  {https://arxiv.org/abs/1905.09687} {arXiv:1905.09687 [astro-ph.CO]}
  \BibitemShut {NoStop}%
\bibitem [{\citenamefont {{Planck Collaboration}}(2016)}]{planck_2015_mg}%
  \BibitemOpen
  \bibfield  {author} {\bibinfo {author} {\bibnamefont {{Planck
  Collaboration}}},\ }\href {https://doi.org/10.1051/0004-6361/201525814}
  {\bibfield  {journal} {\bibinfo  {journal} {\aap}\ }\textbf {\bibinfo
  {volume} {594}},\ \bibinfo {eid} {A14} (\bibinfo {year} {2016})}\BibitemShut
  {NoStop}%
\bibitem [{\citenamefont {Raveri}\ \emph {et~al.}(2021)\citenamefont {Raveri},
  \citenamefont {Pogosian}, \citenamefont {Koyama}, \citenamefont {Martinelli},
  \citenamefont {Silvestri}, \citenamefont {Zhao}, \citenamefont {Li},
  \citenamefont {Peirone},\ and\ \citenamefont {Zucca}}]{raveri2021joint}%
  \BibitemOpen
  \bibfield  {author} {\bibinfo {author} {\bibfnamefont {M.}~\bibnamefont
  {Raveri}}, \bibinfo {author} {\bibfnamefont {L.}~\bibnamefont {Pogosian}},
  \bibinfo {author} {\bibfnamefont {K.}~\bibnamefont {Koyama}}, \bibinfo
  {author} {\bibfnamefont {M.}~\bibnamefont {Martinelli}}, \bibinfo {author}
  {\bibfnamefont {A.}~\bibnamefont {Silvestri}}, \bibinfo {author}
  {\bibfnamefont {G.-B.}\ \bibnamefont {Zhao}}, \bibinfo {author}
  {\bibfnamefont {J.}~\bibnamefont {Li}}, \bibinfo {author} {\bibfnamefont
  {S.}~\bibnamefont {Peirone}},\ and\ \bibinfo {author} {\bibfnamefont
  {A.}~\bibnamefont {Zucca}},\ }\href@noop {} {\bibinfo {title} {A joint
  reconstruction of dark energy and modified growth evolution}} (\bibinfo
  {year} {2021}),\ \Eprint {https://arxiv.org/abs/2107.12990} {arXiv:2107.12990
  [astro-ph.CO]} \BibitemShut {NoStop}%
\bibitem [{\citenamefont {Linder}\ and\ \citenamefont
  {Huterer}(2005)}]{Linder_2005}%
  \BibitemOpen
  \bibfield  {author} {\bibinfo {author} {\bibfnamefont {E.~V.}\ \bibnamefont
  {Linder}}\ and\ \bibinfo {author} {\bibfnamefont {D.}~\bibnamefont
  {Huterer}},\ }\href {https://doi.org/10.1103/PhysRevD.72.043509} {\bibfield
  {journal} {\bibinfo  {journal} {Phys. Rev. D}\ }\textbf {\bibinfo {volume}
  {72}},\ \bibinfo {pages} {043509} (\bibinfo {year} {2005})}\BibitemShut
  {NoStop}%
\bibitem [{\citenamefont {Asaba}\ \emph {et~al.}(2013)\citenamefont {Asaba},
  \citenamefont {Hikage}, \citenamefont {Koyama}, \citenamefont {Zhao},
  \citenamefont {Hojjati},\ and\ \citenamefont {Pogosian}}]{pca_sigmamu}%
  \BibitemOpen
  \bibfield  {author} {\bibinfo {author} {\bibfnamefont {S.}~\bibnamefont
  {Asaba}}, \bibinfo {author} {\bibfnamefont {C.}~\bibnamefont {Hikage}},
  \bibinfo {author} {\bibfnamefont {K.}~\bibnamefont {Koyama}}, \bibinfo
  {author} {\bibfnamefont {G.-B.}\ \bibnamefont {Zhao}}, \bibinfo {author}
  {\bibfnamefont {A.}~\bibnamefont {Hojjati}},\ and\ \bibinfo {author}
  {\bibfnamefont {L.}~\bibnamefont {Pogosian}},\ }\href
  {https://doi.org/10.1088/1475-7516/2013/08/029} {\bibfield  {journal}
  {\bibinfo  {journal} {Journal of Cosmology and Astroparticle Physics}\
  }\textbf {\bibinfo {volume} {2013}}\bibinfo  {number} { (08)},\ \bibinfo
  {pages} {029–029}}\BibitemShut {NoStop}%
\bibitem [{\citenamefont {Kohonen}(1982)}]{kohonen_self-organized_1982}%
  \BibitemOpen
\bibfield  {number} {  }\bibfield  {author} {\bibinfo {author} {\bibfnamefont
  {T.}~\bibnamefont {Kohonen}},\ }\href {https://doi.org/10.1007/BF00337288}
  {\bibfield  {journal} {\bibinfo  {journal} {Biological Cybernetics}\ }\textbf
  {\bibinfo {volume} {43}},\ \bibinfo {pages} {59} (\bibinfo {year}
  {1982})}\BibitemShut {NoStop}%
\bibitem [{\citenamefont {Maehoenen}\ and\ \citenamefont
  {Hakala}(1995)}]{maehoenen_automated_1995}%
  \BibitemOpen
  \bibfield  {author} {\bibinfo {author} {\bibfnamefont {P.~H.}\ \bibnamefont
  {Maehoenen}}\ and\ \bibinfo {author} {\bibfnamefont {P.~J.}\ \bibnamefont
  {Hakala}},\ }\href {https://doi.org/10.1086/309697} {\bibfield  {journal}
  {\bibinfo  {journal} {The Astrophysical Journal Letters}\ }\textbf {\bibinfo
  {volume} {452}},\ \bibinfo {pages} {L77} (\bibinfo {year}
  {1995})}\BibitemShut {NoStop}%
\bibitem [{\citenamefont {Hemmati}\ \emph {et~al.}(2019)\citenamefont
  {Hemmati}, \citenamefont {Capak}, \citenamefont {Pourrahmani}, \citenamefont
  {Nayyeri}, \citenamefont {Stern}, \citenamefont {Mobasher}, \citenamefont
  {Darvish}, \citenamefont {Davidzon}, \citenamefont {Ilbert}, \citenamefont
  {Masters},\ and\ \citenamefont {Shahidi}}]{hemmati_bringing_2019}%
  \BibitemOpen
  \bibfield  {author} {\bibinfo {author} {\bibfnamefont {S.}~\bibnamefont
  {Hemmati}}, \bibinfo {author} {\bibfnamefont {P.}~\bibnamefont {Capak}},
  \bibinfo {author} {\bibfnamefont {M.}~\bibnamefont {Pourrahmani}}, \bibinfo
  {author} {\bibfnamefont {H.}~\bibnamefont {Nayyeri}}, \bibinfo {author}
  {\bibfnamefont {D.}~\bibnamefont {Stern}}, \bibinfo {author} {\bibfnamefont
  {B.}~\bibnamefont {Mobasher}}, \bibinfo {author} {\bibfnamefont
  {B.}~\bibnamefont {Darvish}}, \bibinfo {author} {\bibfnamefont
  {I.}~\bibnamefont {Davidzon}}, \bibinfo {author} {\bibfnamefont
  {O.}~\bibnamefont {Ilbert}}, \bibinfo {author} {\bibfnamefont
  {D.}~\bibnamefont {Masters}},\ and\ \bibinfo {author} {\bibfnamefont
  {A.}~\bibnamefont {Shahidi}},\ }\href
  {https://doi.org/10.3847/2041-8213/ab3418} {\bibfield  {journal} {\bibinfo
  {journal} {The Astrophysical Journal}\ }\textbf {\bibinfo {volume} {881}},\
  \bibinfo {pages} {L14} (\bibinfo {year} {2019})}\BibitemShut {NoStop}%
\bibitem [{\citenamefont {Masters}\ \emph {et~al.}(2015)\citenamefont
  {Masters}, \citenamefont {Capak}, \citenamefont {Stern}, \citenamefont
  {Ilbert}, \citenamefont {Salvato}, \citenamefont {Schmidt}, \citenamefont
  {Longo}, \citenamefont {Rhodes}, \citenamefont {Paltani}, \citenamefont
  {Mobasher} \emph {et~al.}}]{masters_mapping_2015}%
  \BibitemOpen
  \bibfield  {author} {\bibinfo {author} {\bibfnamefont {D.}~\bibnamefont
  {Masters}}, \bibinfo {author} {\bibfnamefont {P.}~\bibnamefont {Capak}},
  \bibinfo {author} {\bibfnamefont {D.}~\bibnamefont {Stern}}, \bibinfo
  {author} {\bibfnamefont {O.}~\bibnamefont {Ilbert}}, \bibinfo {author}
  {\bibfnamefont {M.}~\bibnamefont {Salvato}}, \bibinfo {author} {\bibfnamefont
  {S.}~\bibnamefont {Schmidt}}, \bibinfo {author} {\bibfnamefont
  {G.}~\bibnamefont {Longo}}, \bibinfo {author} {\bibfnamefont
  {J.}~\bibnamefont {Rhodes}}, \bibinfo {author} {\bibfnamefont
  {S.}~\bibnamefont {Paltani}}, \bibinfo {author} {\bibfnamefont
  {B.}~\bibnamefont {Mobasher}}, \emph {et~al.},\ }\href
  {https://doi.org/10.1088/0004-637X/813/1/53} {\bibfield  {journal} {\bibinfo
  {journal} {The Astrophysical Journal}\ }\textbf {\bibinfo {volume} {813}},\
  \bibinfo {pages} {53} (\bibinfo {year} {2015})}\BibitemShut {NoStop}%
\bibitem [{\citenamefont {Wright}\ \emph {et~al.}(2020)\citenamefont {Wright},
  \citenamefont {Hildebrandt}, \citenamefont {van~den Busch},\ and\
  \citenamefont {Heymans}}]{som_photoz1}%
  \BibitemOpen
  \bibfield  {author} {\bibinfo {author} {\bibfnamefont {A.~H.}\ \bibnamefont
  {Wright}}, \bibinfo {author} {\bibfnamefont {H.}~\bibnamefont {Hildebrandt}},
  \bibinfo {author} {\bibfnamefont {J.~L.}\ \bibnamefont {van~den Busch}},\
  and\ \bibinfo {author} {\bibfnamefont {C.}~\bibnamefont {Heymans}},\ }\href
  {https://doi.org/10.1051/0004-6361/201936782} {\bibfield  {journal} {\bibinfo
   {journal} {Astronomy \& Astrophysics}\ }\textbf {\bibinfo {volume} {637}},\
  \bibinfo {pages} {A100} (\bibinfo {year} {2020})}\BibitemShut {NoStop}%
\bibitem [{\citenamefont {Myles}\ \emph {et~al.}(2021)\citenamefont {Myles},
  \citenamefont {Alarcon}, \citenamefont {Amon}, \citenamefont {Sánchez},
  \citenamefont {Everett}, \citenamefont {DeRose}, \citenamefont {McCullough},
  \citenamefont {Gruen}, \citenamefont {Bernstein}, \citenamefont {Troxel}
  \emph {et~al.}}]{som_photoz2}%
  \BibitemOpen
  \bibfield  {author} {\bibinfo {author} {\bibfnamefont {J.}~\bibnamefont
  {Myles}}, \bibinfo {author} {\bibfnamefont {A.}~\bibnamefont {Alarcon}},
  \bibinfo {author} {\bibfnamefont {A.}~\bibnamefont {Amon}}, \bibinfo {author}
  {\bibfnamefont {C.}~\bibnamefont {Sánchez}}, \bibinfo {author}
  {\bibfnamefont {S.}~\bibnamefont {Everett}}, \bibinfo {author} {\bibfnamefont
  {J.}~\bibnamefont {DeRose}}, \bibinfo {author} {\bibfnamefont
  {J.}~\bibnamefont {McCullough}}, \bibinfo {author} {\bibfnamefont
  {D.}~\bibnamefont {Gruen}}, \bibinfo {author} {\bibfnamefont {G.~M.}\
  \bibnamefont {Bernstein}}, \bibinfo {author} {\bibfnamefont {M.~A.}\
  \bibnamefont {Troxel}}, \emph {et~al.},\ }\href
  {https://doi.org/10.1093/mnras/stab1515} {\bibfield  {journal} {\bibinfo
  {journal} {Monthly Notices of the Royal Astronomical Society}\ }\textbf
  {\bibinfo {volume} {505}},\ \bibinfo {pages} {4249} (\bibinfo {year}
  {2021})}\BibitemShut {NoStop}%
\bibitem [{\citenamefont {Moosavi}\ \emph {et~al.}(2014)\citenamefont
  {Moosavi}, \citenamefont {Packmann},\ and\ \citenamefont
  {Vall{\'e}s}}]{moosavi2014sompy}%
  \BibitemOpen
  \bibfield  {author} {\bibinfo {author} {\bibfnamefont {V.}~\bibnamefont
  {Moosavi}}, \bibinfo {author} {\bibfnamefont {S.}~\bibnamefont {Packmann}},\
  and\ \bibinfo {author} {\bibfnamefont {I.}~\bibnamefont {Vall{\'e}s}},\
  }\href@noop {} {\bibinfo {title} {Sompy: A python library for self organizing
  map (som)}} (\bibinfo {year} {2014}),\ \bibinfo {note} {gitHub.[Online].
  Available: https://github. com/sevamoo/SOMPY}\BibitemShut {NoStop}%
\bibitem [{\citenamefont {Davidzon}\ \emph {et~al.}(2019)\citenamefont
  {Davidzon}, \citenamefont {Laigle}, \citenamefont {Capak}, \citenamefont
  {Ilbert}, \citenamefont {Masters}, \citenamefont {Hemmati}, \citenamefont
  {Apostolakos}, \citenamefont {Coupon}, \citenamefont {de~la Torre},
  \citenamefont {Devriendt}, \citenamefont {Dubois}, \citenamefont {Kashino},
  \citenamefont {Paltani},\ and\ \citenamefont
  {Pichon}}]{davidzon_horizon-agn_2019}%
  \BibitemOpen
  \bibfield  {author} {\bibinfo {author} {\bibfnamefont {I.}~\bibnamefont
  {Davidzon}}, \bibinfo {author} {\bibfnamefont {C.}~\bibnamefont {Laigle}},
  \bibinfo {author} {\bibfnamefont {P.~L.}\ \bibnamefont {Capak}}, \bibinfo
  {author} {\bibfnamefont {O.}~\bibnamefont {Ilbert}}, \bibinfo {author}
  {\bibfnamefont {D.~C.}\ \bibnamefont {Masters}}, \bibinfo {author}
  {\bibfnamefont {S.}~\bibnamefont {Hemmati}}, \bibinfo {author} {\bibfnamefont
  {N.}~\bibnamefont {Apostolakos}}, \bibinfo {author} {\bibfnamefont
  {J.}~\bibnamefont {Coupon}}, \bibinfo {author} {\bibfnamefont
  {S.}~\bibnamefont {de~la Torre}}, \bibinfo {author} {\bibfnamefont
  {J.}~\bibnamefont {Devriendt}}, \bibinfo {author} {\bibfnamefont
  {Y.}~\bibnamefont {Dubois}}, \bibinfo {author} {\bibfnamefont
  {D.}~\bibnamefont {Kashino}}, \bibinfo {author} {\bibfnamefont
  {S.}~\bibnamefont {Paltani}},\ and\ \bibinfo {author} {\bibfnamefont
  {C.}~\bibnamefont {Pichon}},\ }\href {https://doi.org/10.1093/mnras/stz2486}
  {\bibfield  {journal} {\bibinfo  {journal} {Monthly Notices of the Royal
  Astronomical Society}\ }\textbf {\bibinfo {volume} {489}},\ \bibinfo {pages}
  {4817} (\bibinfo {year} {2019})}\BibitemShut {NoStop}%
\bibitem [{\citenamefont {Zuntz}\ \emph {et~al.}(2015)\citenamefont {Zuntz},
  \citenamefont {Paterno}, \citenamefont {Jennings}, \citenamefont {Rudd},
  \citenamefont {Manzotti}, \citenamefont {Dodelson}, \citenamefont {Bridle},
  \citenamefont {Sehrish},\ and\ \citenamefont
  {Kowalkowski}}]{zuntz_cosmosis_2015}%
  \BibitemOpen
  \bibfield  {author} {\bibinfo {author} {\bibfnamefont {J.}~\bibnamefont
  {Zuntz}}, \bibinfo {author} {\bibfnamefont {M.}~\bibnamefont {Paterno}},
  \bibinfo {author} {\bibfnamefont {E.}~\bibnamefont {Jennings}}, \bibinfo
  {author} {\bibfnamefont {D.}~\bibnamefont {Rudd}}, \bibinfo {author}
  {\bibfnamefont {A.}~\bibnamefont {Manzotti}}, \bibinfo {author}
  {\bibfnamefont {S.}~\bibnamefont {Dodelson}}, \bibinfo {author}
  {\bibfnamefont {S.}~\bibnamefont {Bridle}}, \bibinfo {author} {\bibfnamefont
  {S.}~\bibnamefont {Sehrish}},\ and\ \bibinfo {author} {\bibfnamefont
  {J.}~\bibnamefont {Kowalkowski}},\ }\href
  {https://doi.org/10.1016/j.ascom.2015.05.005} {\bibfield  {journal} {\bibinfo
   {journal} {Astronomy and Computing}\ }\textbf {\bibinfo {volume} {12}},\
  \bibinfo {pages} {45} (\bibinfo {year} {2015})}\BibitemShut {NoStop}%
\bibitem [{\citenamefont {Zucca}\ \emph {et~al.}(2019)\citenamefont {Zucca},
  \citenamefont {Pogosian}, \citenamefont {Silvestri},\ and\ \citenamefont
  {Zhao}}]{zucca_mgcamb_2019}%
  \BibitemOpen
  \bibfield  {author} {\bibinfo {author} {\bibfnamefont {A.}~\bibnamefont
  {Zucca}}, \bibinfo {author} {\bibfnamefont {L.}~\bibnamefont {Pogosian}},
  \bibinfo {author} {\bibfnamefont {A.}~\bibnamefont {Silvestri}},\ and\
  \bibinfo {author} {\bibfnamefont {G.-B.}\ \bibnamefont {Zhao}},\ }\href
  {https://doi.org/10.1088/1475-7516/2019/05/001} {\bibfield  {journal}
  {\bibinfo  {journal} {Journal of Cosmology and Astroparticle Physics}\
  }\textbf {\bibinfo {volume} {2019}}\bibinfo  {number} { (05)},\ \bibinfo
  {pages} {001}}\BibitemShut {NoStop}%
\bibitem [{\citenamefont {Hojjati}\ \emph {et~al.}(2011)\citenamefont
  {Hojjati}, \citenamefont {Pogosian},\ and\ \citenamefont
  {Zhao}}]{hojjati_testing_2011}%
  \BibitemOpen
\bibfield  {number} {  }\bibfield  {author} {\bibinfo {author} {\bibfnamefont
  {A.}~\bibnamefont {Hojjati}}, \bibinfo {author} {\bibfnamefont
  {L.}~\bibnamefont {Pogosian}},\ and\ \bibinfo {author} {\bibfnamefont
  {G.-B.}\ \bibnamefont {Zhao}},\ }\href
  {https://doi.org/10.1088/1475-7516/2011/08/005} {\bibfield  {journal}
  {\bibinfo  {journal} {Journal of Cosmology and Astroparticle Physics}\
  }\textbf {\bibinfo {volume} {2011}}\bibinfo  {number} { (08)},\ \bibinfo
  {pages} {005}}\BibitemShut {NoStop}%
\bibitem [{\citenamefont {Zhao}\ \emph {et~al.}(2009)\citenamefont {Zhao},
  \citenamefont {Pogosian}, \citenamefont {Silvestri},\ and\ \citenamefont
  {Zylberberg}}]{zhao_searching_2009}%
  \BibitemOpen
\bibfield  {number} {  }\bibfield  {author} {\bibinfo {author} {\bibfnamefont
  {G.-B.}\ \bibnamefont {Zhao}}, \bibinfo {author} {\bibfnamefont
  {L.}~\bibnamefont {Pogosian}}, \bibinfo {author} {\bibfnamefont
  {A.}~\bibnamefont {Silvestri}},\ and\ \bibinfo {author} {\bibfnamefont
  {J.}~\bibnamefont {Zylberberg}},\ }\href
  {https://doi.org/10.1103/PhysRevD.79.083513} {\bibfield  {journal} {\bibinfo
  {journal} {Physical Review D}\ }\textbf {\bibinfo {volume} {79}},\ \bibinfo
  {pages} {083513} (\bibinfo {year} {2009})}\BibitemShut {NoStop}%
\bibitem [{\citenamefont {Hu}\ and\ \citenamefont {Sawicki}(2007)}]{Hu_2007}%
  \BibitemOpen
  \bibfield  {author} {\bibinfo {author} {\bibfnamefont {W.}~\bibnamefont
  {Hu}}\ and\ \bibinfo {author} {\bibfnamefont {I.}~\bibnamefont {Sawicki}},\
  }\href {https://doi.org/10.1103/PhysRevD.76.064004} {\bibfield  {journal}
  {\bibinfo  {journal} {Phys. Rev. D}\ }\textbf {\bibinfo {volume} {76}},\
  \bibinfo {pages} {064004} (\bibinfo {year} {2007})}\BibitemShut {NoStop}%
\bibitem [{\citenamefont {Song}\ \emph {et~al.}(2007)\citenamefont {Song},
  \citenamefont {Hu},\ and\ \citenamefont {Sawicki}}]{fr_b0}%
  \BibitemOpen
  \bibfield  {author} {\bibinfo {author} {\bibfnamefont {Y.-S.}\ \bibnamefont
  {Song}}, \bibinfo {author} {\bibfnamefont {W.}~\bibnamefont {Hu}},\ and\
  \bibinfo {author} {\bibfnamefont {I.}~\bibnamefont {Sawicki}},\ }\href
  {https://doi.org/10.1103/PhysRevD.75.044004} {\bibfield  {journal} {\bibinfo
  {journal} {Phys. Rev. D}\ }\textbf {\bibinfo {volume} {75}},\ \bibinfo
  {pages} {044004} (\bibinfo {year} {2007})}\BibitemShut {NoStop}%
\bibitem [{\citenamefont {Dossett}\ \emph {et~al.}(2014)\citenamefont
  {Dossett}, \citenamefont {Hu},\ and\ \citenamefont
  {Parkinson}}]{dossett_constraining_2014}%
  \BibitemOpen
  \bibfield  {author} {\bibinfo {author} {\bibfnamefont {J.}~\bibnamefont
  {Dossett}}, \bibinfo {author} {\bibfnamefont {B.}~\bibnamefont {Hu}},\ and\
  \bibinfo {author} {\bibfnamefont {D.}~\bibnamefont {Parkinson}},\ }\href
  {https://doi.org/10.1088/1475-7516/2014/03/046} {\bibfield  {journal}
  {\bibinfo  {journal} {Journal of Cosmology and Astroparticle Physics}\
  }\textbf {\bibinfo {volume} {2014}}\bibinfo  {number} { (03)},\ \bibinfo
  {pages} {046}}\BibitemShut {NoStop}%
\bibitem [{\citenamefont {Damour}\ and\ \citenamefont
  {Polyakov}(1994)}]{dilaton_screen}%
  \BibitemOpen
\bibfield  {number} {  }\bibfield  {author} {\bibinfo {author} {\bibfnamefont
  {T.}~\bibnamefont {Damour}}\ and\ \bibinfo {author} {\bibfnamefont
  {A.}~\bibnamefont {Polyakov}},\ }\href
  {https://doi.org/10.1016/0550-3213(94)90143-0} {\bibfield  {journal}
  {\bibinfo  {journal} {Nuclear Physics B}\ }\textbf {\bibinfo {volume}
  {423}},\ \bibinfo {pages} {532–558} (\bibinfo {year} {1994})}\BibitemShut
  {NoStop}%
\bibitem [{\citenamefont {Brax}\ \emph {et~al.}(2010)\citenamefont {Brax},
  \citenamefont {van~de Bruck}, \citenamefont {Davis},\ and\ \citenamefont
  {Shaw}}]{dilaton_cosmology}%
  \BibitemOpen
  \bibfield  {author} {\bibinfo {author} {\bibfnamefont {P.}~\bibnamefont
  {Brax}}, \bibinfo {author} {\bibfnamefont {C.}~\bibnamefont {van~de Bruck}},
  \bibinfo {author} {\bibfnamefont {A.-C.}\ \bibnamefont {Davis}},\ and\
  \bibinfo {author} {\bibfnamefont {D.}~\bibnamefont {Shaw}},\ }\href
  {https://doi.org/10.1103/PhysRevD.82.063519} {\bibfield  {journal} {\bibinfo
  {journal} {Phys. Rev. D}\ }\textbf {\bibinfo {volume} {82}},\ \bibinfo
  {pages} {063519} (\bibinfo {year} {2010})}\BibitemShut {NoStop}%
\bibitem [{\citenamefont {Hinterbichler}\ and\ \citenamefont
  {Khoury}(2010)}]{symmetron_cosmology}%
  \BibitemOpen
  \bibfield  {author} {\bibinfo {author} {\bibfnamefont {K.}~\bibnamefont
  {Hinterbichler}}\ and\ \bibinfo {author} {\bibfnamefont {J.}~\bibnamefont
  {Khoury}},\ }\href {https://doi.org/10.1103/PhysRevLett.104.231301}
  {\bibfield  {journal} {\bibinfo  {journal} {Phys. Rev. Lett.}\ }\textbf
  {\bibinfo {volume} {104}},\ \bibinfo {pages} {231301} (\bibinfo {year}
  {2010})}\BibitemShut {NoStop}%
\bibitem [{\citenamefont {Vazsonyi}\ \emph {et~al.}(2021)\citenamefont
  {Vazsonyi}, \citenamefont {Taylor}, \citenamefont {Valogiannis},
  \citenamefont {Ramachandra}, \citenamefont {Fert\'e},\ and\ \citenamefont
  {Rhodes}}]{Vazsonyi_2021}%
  \BibitemOpen
  \bibfield  {author} {\bibinfo {author} {\bibfnamefont {L.}~\bibnamefont
  {Vazsonyi}}, \bibinfo {author} {\bibfnamefont {P.~L.}\ \bibnamefont
  {Taylor}}, \bibinfo {author} {\bibfnamefont {G.}~\bibnamefont {Valogiannis}},
  \bibinfo {author} {\bibfnamefont {N.~S.}\ \bibnamefont {Ramachandra}},
  \bibinfo {author} {\bibfnamefont {A.}~\bibnamefont {Fert\'e}},\ and\ \bibinfo
  {author} {\bibfnamefont {J.}~\bibnamefont {Rhodes}},\ }\href
  {https://doi.org/10.1103/PhysRevD.104.083527} {\bibfield  {journal} {\bibinfo
   {journal} {Phys. Rev. D}\ }\textbf {\bibinfo {volume} {104}},\ \bibinfo
  {pages} {083527} (\bibinfo {year} {2021})}\BibitemShut {NoStop}%
\bibitem [{\citenamefont {Jain}\ \emph {et~al.}(2013)\citenamefont {Jain},
  \citenamefont {Vikram},\ and\ \citenamefont {Sakstein}}]{Jain_2013}%
  \BibitemOpen
  \bibfield  {author} {\bibinfo {author} {\bibfnamefont {B.}~\bibnamefont
  {Jain}}, \bibinfo {author} {\bibfnamefont {V.}~\bibnamefont {Vikram}},\ and\
  \bibinfo {author} {\bibfnamefont {J.}~\bibnamefont {Sakstein}},\ }\href
  {https://doi.org/10.1088/0004-637x/779/1/39} {\bibfield  {journal} {\bibinfo
  {journal} {The Astrophysical Journal}\ }\textbf {\bibinfo {volume} {779}},\
  \bibinfo {pages} {39} (\bibinfo {year} {2013})}\BibitemShut {NoStop}%
\bibitem [{\citenamefont {Desmond}\ and\ \citenamefont
  {Ferreira}(2020)}]{Desmond_2020}%
  \BibitemOpen
  \bibfield  {author} {\bibinfo {author} {\bibfnamefont {H.}~\bibnamefont
  {Desmond}}\ and\ \bibinfo {author} {\bibfnamefont {P.~G.}\ \bibnamefont
  {Ferreira}},\ }\href {https://doi.org/10.1103/PhysRevD.102.104060} {\bibfield
   {journal} {\bibinfo  {journal} {Phys. Rev. D}\ }\textbf {\bibinfo {volume}
  {102}},\ \bibinfo {pages} {104060} (\bibinfo {year} {2020})}\BibitemShut
  {NoStop}%
\bibitem [{\citenamefont {Krause}\ and\ \citenamefont
  {Eifler}(2017)}]{cosmocov1}%
  \BibitemOpen
  \bibfield  {author} {\bibinfo {author} {\bibfnamefont {E.}~\bibnamefont
  {Krause}}\ and\ \bibinfo {author} {\bibfnamefont {T.}~\bibnamefont
  {Eifler}},\ }\href {https://doi.org/10.1093/mnras/stx1261} {\bibfield
  {journal} {\bibinfo  {journal} {Monthly Notices of the Royal Astronomical
  Society}\ }\textbf {\bibinfo {volume} {470}},\ \bibinfo {pages} {2100–2112}
  (\bibinfo {year} {2017})}\BibitemShut {NoStop}%
\bibitem [{\citenamefont {Fang}\ \emph
  {et~al.}(2020{\natexlab{a}})\citenamefont {Fang}, \citenamefont {Krause},
  \citenamefont {Eifler},\ and\ \citenamefont {MacCrann}}]{cosmocov2}%
  \BibitemOpen
  \bibfield  {author} {\bibinfo {author} {\bibfnamefont {X.}~\bibnamefont
  {Fang}}, \bibinfo {author} {\bibfnamefont {E.}~\bibnamefont {Krause}},
  \bibinfo {author} {\bibfnamefont {T.}~\bibnamefont {Eifler}},\ and\ \bibinfo
  {author} {\bibfnamefont {N.}~\bibnamefont {MacCrann}},\ }\href
  {https://doi.org/10.1088/1475-7516/2020/05/010} {\bibfield  {journal}
  {\bibinfo  {journal} {Journal of Cosmology and Astroparticle Physics}\
  }\textbf {\bibinfo {volume} {2020}}\bibinfo  {number} { (05)},\ \bibinfo
  {pages} {010–010}}\BibitemShut {NoStop}%
\bibitem [{\citenamefont {Fang}\ \emph
  {et~al.}(2020{\natexlab{b}})\citenamefont {Fang}, \citenamefont {Eifler},\
  and\ \citenamefont {Krause}}]{cosmocov3}%
  \BibitemOpen
\bibfield  {number} {  }\bibfield  {author} {\bibinfo {author} {\bibfnamefont
  {X.}~\bibnamefont {Fang}}, \bibinfo {author} {\bibfnamefont {T.}~\bibnamefont
  {Eifler}},\ and\ \bibinfo {author} {\bibfnamefont {E.}~\bibnamefont
  {Krause}},\ }\href {https://doi.org/10.1093/mnras/staa1726} {\bibfield
  {journal} {\bibinfo  {journal} {Monthly Notices of the Royal Astronomical
  Society}\ }\textbf {\bibinfo {volume} {497}},\ \bibinfo {pages} {2699–2714}
  (\bibinfo {year} {2020}{\natexlab{b}})}\BibitemShut {NoStop}%
\bibitem [{\citenamefont {Linder}(2003)}]{w0wa_1}%
  \BibitemOpen
  \bibfield  {author} {\bibinfo {author} {\bibfnamefont {E.~V.}\ \bibnamefont
  {Linder}},\ }\href {https://doi.org/10.1103/PhysRevLett.90.091301} {\bibfield
   {journal} {\bibinfo  {journal} {Phys. Rev. Lett.}\ }\textbf {\bibinfo
  {volume} {90}},\ \bibinfo {pages} {091301} (\bibinfo {year}
  {2003})}\BibitemShut {NoStop}%
\bibitem [{\citenamefont {Chevallier}\ and\ \citenamefont
  {Polarski}(2001)}]{w0wa_2}%
  \BibitemOpen
  \bibfield  {author} {\bibinfo {author} {\bibfnamefont {M.}~\bibnamefont
  {Chevallier}}\ and\ \bibinfo {author} {\bibfnamefont {D.}~\bibnamefont
  {Polarski}},\ }\href {https://doi.org/10.1142/s0218271801000822} {\bibfield
  {journal} {\bibinfo  {journal} {International Journal of Modern Physics D}\
  }\textbf {\bibinfo {volume} {10}},\ \bibinfo {pages} {213–223} (\bibinfo
  {year} {2001})}\BibitemShut {NoStop}%
\bibitem [{\citenamefont {Lewis}\ \emph {et~al.}(2000)\citenamefont {Lewis},
  \citenamefont {Challinor},\ and\ \citenamefont {Lasenby}}]{Lewis:1999bs}%
  \BibitemOpen
  \bibfield  {author} {\bibinfo {author} {\bibfnamefont {A.}~\bibnamefont
  {Lewis}}, \bibinfo {author} {\bibfnamefont {A.}~\bibnamefont {Challinor}},\
  and\ \bibinfo {author} {\bibfnamefont {A.}~\bibnamefont {Lasenby}},\ }\href
  {https://doi.org/10.1086/309179} {\bibfield  {journal} {\bibinfo  {journal}
  {\apj}\ }\textbf {\bibinfo {volume} {538}},\ \bibinfo {pages} {473} (\bibinfo
  {year} {2000})}\BibitemShut {NoStop}%
\bibitem [{\citenamefont {Howlett}\ \emph {et~al.}(2012)\citenamefont
  {Howlett}, \citenamefont {Lewis}, \citenamefont {Hall},\ and\ \citenamefont
  {Challinor}}]{Howlett:2012mh}%
  \BibitemOpen
  \bibfield  {author} {\bibinfo {author} {\bibfnamefont {C.}~\bibnamefont
  {Howlett}}, \bibinfo {author} {\bibfnamefont {A.}~\bibnamefont {Lewis}},
  \bibinfo {author} {\bibfnamefont {A.}~\bibnamefont {Hall}},\ and\ \bibinfo
  {author} {\bibfnamefont {A.}~\bibnamefont {Challinor}},\ }\href
  {https://doi.org/10.1088/1475-7516/2012/04/027} {\bibfield  {journal}
  {\bibinfo  {journal} {Journal of Cosmology and Astroparticle Physics}\
  }\textbf {\bibinfo {volume} {2012}}\bibinfo  {number} { (04)},\ \bibinfo
  {pages} {027}}\BibitemShut {NoStop}%
\bibitem [{\citenamefont {Lewis}(2014)}]{camb_notes}%
  \BibitemOpen
\bibfield  {number} {  }\bibfield  {author} {\bibinfo {author} {\bibfnamefont
  {A.}~\bibnamefont {Lewis}}} (\bibinfo {year} {2014}),\ \bibinfo {note}
  {\url{https://cosmologist.info/notes/CAMB.pdf}}\BibitemShut {NoStop}%
\end{thebibliography}%

\end{document}